\documentclass[12pt,twoside,a4paper]{article}
\usepackage{rawfonts,amsmath,amsfonts,amssymb,epsfig,color}
\catcode`\@=11
\def\QED{\hbox{\hskip 1pt \vrule width4pt height 6pt depth 1.5pt \hskip 1pt}}

\newtheorem{theo}{Theorem}

\newtheorem{lemma}{Lemma}
\def\DOM{{\rm DOM}}
\def\lev{{\rm level}}

\title{Maximum Matching in General Graphs Without Explicit Consideration 
of Blossoms Revisited}

\author{Norbert Blum\thanks{Institut f\"ur Informatik, Universit\"at Bonn, Friedrich-Ebert-Allee 144, D-53113 Bonn, Germany.
    e-mail:{ \tt blum@cs.uni-bonn.de}}}
\begin{document}
\date{}
\maketitle
\begin{abstract}
\noindent 
We reduce the problem of finding an augmenting path in a general
graph to a reachability problem in a directed bipartite graph. 
A slight modification of depth-first search leads to an algorithm for 
finding such paths.
Although this setting is equivalent to the traditional terminology of 
blossoms due to Edmonds, there are some advantages. Mainly, 
this point of view enables the description of algorithms
for the solution of matching problems without explicit analysis of
blossoms, nested blossoms, and so on. Exemplary, we describe an efficient
realization of the Hopcroft-Karp approach for the computation of a 
maximum cardinality matching in general graphs and a variant of 
Edmonds' primal-dual algorithm for the maximum weighted matching problem.
\end{abstract}

\section{Introduction and motivation}

Since Berge's theorem in 1957 \cite{Be} it has been well known that
for constructing a maximum matching, it suffices to search for augmenting
paths. But until 1965, only exponential algorithms for finding a maximum 
cardinality matching in non-bipartite graphs have been known. The reason was that 
one did not know how to treat odd cycles, the so-called ``blossoms'', in 
alternating paths.
In his pioneering work, Edmonds \cite{Ed1} solved this problem by shrinking
these odd cycles. Because each shrinking reduces the number of nodes in the
current graph at least by two, the total number of shrinkage's is bounded by
$\frac{n}{2}$ where $n$ is the number of nodes in the graphs. Hence, Edmonds
original algorithm uses only $O(n^3)$ time.
In \cite{Bal,Gab,La,WZ}, it is shown how to avoid explicit
shrinking of odd cycles. All these algorithms use $O(n^3)$ or $O(nm\alpha(m+n,n))$ 
time where $m$ is the number of edges in the graph and $\alpha$ is the functional 
inverse of Ackermann's function.

In 1973, Hopcroft and Karp \cite{HK} proved the following fact. If one computes
in one phase a maximal set of pairwise disjoint shortest augmenting paths and augments 
these paths then $O(\sqrt{n})$ such
phases would be sufficient. For the bipartite case they showed that a phase can
be implemented using a breadth-first search followed by a depth-first search.
This led to an $O(n+m)$ implementation of one phase and hence, to an $O(\sqrt{n}m)$
algorithm for the computation of a maximum matching in bipartite graphs.
The implementation of a phase for non-bipartite graphs is much harder.
In 1975, Even and Kariv \cite{EK,Ka} presented an $O(\min\{n^2,m\log n\})$
implementation of a phase leading to an $O(\min\{n^{2.5},\sqrt{n}m\log n\})$ 
algorithm for the computation of a maximum matching in general graphs.
Galil \cite{Gal} called the full paper \cite{Ka} ``a strong contender for the ACM 
Longest Paper Award''. Tarjan \cite{Ta} called their paper ``a remarkable 
tour-de-force''.
In 1978, Bartnik \cite{Bar} gave an alternative $O(n^2)$ implementation
in his unpublished Ph.D.~thesis (see \cite{GoMi}).
In 1980, Micali and Vijay Vazirani \cite{MV} have presented an $O(m+n)$
implementation of a phase without the presentation of a correctness proof. 
The algorithm as presented in \cite{MV} is not correct since their definition
of ``tenacity'' does not work in all cases. In 1988, Peterson and Loui
have given an informal correctness proof of the incorrect algorithm of Micali
and Vazirani. Fourteen years later, Vijay Vazirani \cite{Va1} has repaired the 
mistake by changing the definition of ``tenacity'' and provided a correctness proof. 
Recently \cite{Va2}, Vazirani has presented a new version of a correctness proof.
With respect to his proof in \cite{Va1}, he writes: ``Although the statements of these theorems
were largely correct, their proofs, which involved low level arguments about individual
paths and their complicated intersections with other structures, were, in retrospect,
incorrect.'' In 1999 \cite{Bl3}, I have tried to
combine ideas of Micali and Vazirani and my framework. As pointed out by Ross McConnell
\cite{Mc}, the algorithm, as described in \cite{Bl3}, is not correct. In 1991,
Gabow and Tarjan \cite{GaTa2} have given an efficient scaling algorithm for general
graph-matching problems. As pointed out in their paper, by a slight modification of
their algorithm, they have obtained another $O(m+n)$ implementation of a phase.
The history of efficient implementations of a phase of the Hopcroft-Karp approach
for general graphs illustrates the need of a framework which allows a clear description 
and an elaborated correctness proof of matching algorithms.

The first polynomial algorithm for the maximum weighted matching problem has been also 
given by Edmonds \cite{Ed2}. A straightforward implementation of his algorithm has run 
time $O(n^2m)$. Gabow \cite{Ga1} and Lawler \cite{La} have developed $O(n^3)$ 
implementations of Edmonds' algorithm. Galil, Micali and Gabow \cite{GaMiGa} have given 
an $O(nm \log n)$ implementation. At the first SODA, Gabow \cite{Ga2} has
presented an implementation of Edmonds algorithm which uses complicated data structures
and stated that the time complexity of his implementation is $O(n(m + n \log n))$. 

Our goal is to avoid sophisticated explicit analysis of (nested) blossoms.
Hence, we reduce the problem of finding an augmenting path to a reachability 
problem in a directed, bipartite graph. We show, how to solve this reachability 
problem by a modified depth-first search. This approach yields an algorithm which is not
fundamentally different from previous algorithms which use Edmonds' traditional
terminology of blossoms. But if Edmonds' algorithm is used as a subroutine with respect
to the solution of a more involved problem, using the framework of the reachability
problem which avoids the explicit consideration of blossoms can simplify the situation 
considerably. To illustrate this, we describe a realization of the Hopcroft-Karp approach 
\cite{HK} for the 
computation of a maximum cardinality matching in general graphs. Furthermore, we show how 
to use the modified depth-first search algorithm in the primal step of Edmonds' maximum 
weighted matching algorithm. For the description of the primal-dual method, we use no 
linear program and no duality theory for linear programs. 
A straightforward $O(nm \log n)$ implementation will be described as well.

In Section~2, the basic algorithm is presented. After the description of
the reduction to a reachability problem in a directed, bipartite graph, this 
reachability problem is solved by a modification of depth-first search.
The correctness of the algorithm is proved and an efficient implementation
is given. In Section 3, a realization of the Hopcroft-Karp approach for general graphs
is described and its correctness is proved using the framework of the reachability 
problem. Furthermore, an efficient implementation is given.
In Section 4, we show how to use the modified depth-first search as a subroutine in Edmonds' 
maximum weighted matching algorithm. Furthermore, we describe an $O(nm \log n)$
implementation of this approach.

\section{The basic algorithm}

After the reduction of the problem of finding an augmenting path to a reachability
problem, we shall describe the solution of the reachability problem by a modification
of depth-first search. The resulting algorithm is equivalent to Edmonds' algorithm
since to each run of Edmonds' algorithm there corresponds a run of the modified
depth-first search and vice versa \cite{Ro}. Hence, the correctness of the modified
depth-first search follows directly from the correctness of Edmonds' algorithm.
Nevertheless, we shall prove the correctness of the algorithm directly without
the use of the correctness of Edmonds' algorithm. Then we shall describe an efficient
implementation of the algorithm. The description of Edmonds' algorithm within
the framework of the reachability problem seems not to be simpler than an
elaborated description of Edmonds' algorithm within the traditional framework
of blossoms. However, as we shall see later, if Edmonds' algorithm is used as a 
subroutine with respect to the solution of a more involved problem, using the framework
of the reachability problem can simplify the situation considerably.

\subsection{Definitions and the general method}

A {\em graph} $G=(V,E)$ consists of a finite, nonempty set of {\em nodes\/}
$V$ and a set of {\em edges\/} $E$. $G$ is either {\em directed\/} or
{\em undirected\/}. In the (un-)directed case, each edge is an (un-)ordered
pair of distinct nodes. 
A graph $G=(V,E)$ is {\em bipartite\/} if $V$ can be
partitioned into disjoint nonempty sets $A$ and $B$ such that for all
$(u,v)\in E$, $u\in A$ and $v\in B$ or vice versa. Then we often write 
$G=(A,B,E)$.
A {\em path\/} $P$ from $v\in V$ to $w\in V$ is a sequence of
nodes $v=v_0, v_1, \ldots,v_k=w$, which satisfies $(v_i,v_{i+1})\in E$, for
$0\leq i < k$.
The {\em length\/} $|P|$ of $P$ is the number $k$ of edges on $P$. $P$ is
{\em simple\/} if $v_i\not= v_j$, for $0\leq i<j\leq k$. For conveniences,
$P$ will denote the path $v_0, v_1,\ldots,v_k$, the set of nodes
$\{v_0,v_1,\ldots,v_k\}$, and the set of edges $\{(v_0,v_1),(v_1,v_2),
\ldots,(v_{k-1},v_k)\}$.
If there exists a path from $v$ to $w$ (of length $1$) then $v$ is called a
{\em (direct) predecessor\/} of $w$, and $w$ is called a {\em (direct) 
successor\/} of $v$.
Let $G =(V,E)$ be an undirected graph. $M\subseteq E$ is a
{\em matching\/} of $G$ if no two edges in $M$ have a node in common.
A matching $M$ is {\em maximal\/} if there exists no $e\in E\setminus M$ such
that $M\cup \{e\}$ is a matching. A matching $M$ is {\em maximum\/} if there
exists no matching $M'\subseteq E$ of larger size.
Given an undirected graph $G=(V,E)$, the {\em maximum matching problem\/} is
finding a maximum matching $M\subseteq E$.
A path $P=v_0,v_1,\ldots, v_k$ is {\em M-alternating\/}, if it contains 
alternately edges in $M$ and in $E \setminus M$.
A node $v\in V$ is {\em M-free\/} if $v$ is not incident to any edge in $M$. 
Let $P=v_0,v_1,\ldots,v_k$ be a simple $M$-alternating path. $P$ is
{\em $M$-augmenting} if $v_0$ and $v_k$ are $M$-free.
Let $P$ be an $M$-augmenting path in $G$. Then $M\oplus P$ denotes the
{\em symmetric difference\/} of $M$ and $P$; i.e., $M\oplus P = M\setminus  P
\cup  P\setminus M$.
It is easy to see that $M\oplus P$ is a matching of $G$, and $|M\oplus P| =
|M| + 1$.

The key to most algorithms for finding a maximum matching in a graph is the
following theorem of Berge \cite{Be}.

\begin{theo}
Let $G =(V,E)$ be an undirected graph and $M\subseteq E$ be a matching. Then
$M$ is maximum if and only if there exists no $M$-augmenting path in $G$.
\end{theo}
Berge's theorem directly implies the following general method for
finding a maximum matching in a graph $G$.

\medskip
\noindent
{\bf Algorithm 1} \\
\noindent {\bf Input:} An undirected graph $G=(V,E)$, and a matching
$M \subseteq E$. \\
\noindent {\bf Output:} A maximum matching $M_{max}$. \\
\noindent {\bf Method:}
\vspace{-0.3cm}
\begin{tabbing}
AA \= AA \= \kill
{\bf while} there exists an $M$-augmenting path \\
\> {\bf do} \\
        \>  \> construct such a path $P$; \\
        \>  \> $M:= M\oplus P$ \\
\> {\bf od}; \\
$M_{max}$ := $M$.
\end{tabbing}

The key problem is now this: How to find an $M$-augmenting path $P$, if such 
a path exists?
We solve this key problem in the following way.
\begin{enumerate}
\item We reduce the key problem to a reachability problem in a directed,
bipartite graph $G_M = (V',E_M)$.
\item We solve this reachability problem constructively.
\end{enumerate}

\subsection{The reduction to a reachability problem}

In the bipartite case, we construct from $G=(A,B,E)$ and a matching 
$M\subseteq E$ a directed graph $G_M=(A \cup \{s\}, B \cup \{t\},E_M)$ by directing 
the edges in $M$ from~$A$ to~$B$,
and directing the edges in $E\setminus M$ from $B$~to~$A$. Additionally, we add
two new
nodes~$s$ and $t$ to~$A\cup B$, add for each $M$-free node $b\in B$ the edge
$(s,b)$ to~$E_M$, and add for each $M$-free node~$a\in A$ the edge~$(a,t)$
to~$E_M$.
It is easy to prove that there is an $M$-augmenting path in $G$ if and
only if there is a simple path from~$s$ to~$t$ in $G_M$.
This reachability problem can be solved by performing a depth-first search 
of $G_M$ with start node $s$. Now we shall consider the general case.

Let $G=(V,E)$ be an undirected graph and $M\subseteq E$ be a matching. Let
$V_M := \{x\in V \mid x \mbox{ is $M$-free}\}$. For the definition of
$G_M$ we have the following difficulty. Let us consider the graph described in
Figure 1 where edges in $M$ are wavy. The $M$-augmenting path from $x$ to $y$ enters 
the edge $e$ in $b$ and
leaves the edge from $a$. The $M$-augmenting path from $v$ to $w$ enters $e$
in $a$ and leaves $e$ from $b$.
\begin{figure}[t]
\input{nmbild1.pspdftex}
\caption{Difficulty with respect to the definition of $G_M$.}
\end{figure}
A priori, we cannot divide the set of nodes $V$ into two sets $A$ and $B$ such
that an $M$-augmenting path exists in $G$ if and
only if there exists an $M$-augmenting path, using alternately nodes
from $A$ and from $B$. 
Hence, for defining $G_M$, we introduce for each node
$v\in V$ two nodes $[v,A]$ and $[v,B]$ such that an analogous construction of 
a graph $G_M$ is possible.
Both edges $([v,A],[w,B])$ and $([w,A],[v,B])$ are in $G_M$ if and only if
$(v,w) \in M$. Both edges $([x,B],[y,A])$ and $([y,B],[x,A])$ 
are in $G_M$ if and only if $(x,y) \in E \setminus M$. Additionally, we add
for each $M$-free node $v \in V$ the edges $(s,[v,B])$ and $([v,A],t)$ to
$G_M$, where $s$ and $t$ are two new distinct nodes.
More formally, let $G_M := (V',E_M)$ where
\begin{eqnarray*}
V' &:=& \left\{[v,A],[v,B] \mid v\in V\right\}
       \cup \{s,t\} \qquad s,t\not\in V,\ s\not=t \\
E_M &:=& \left\{([v,A],[w,B]),([w,A],[v,B]) \mid (v,w)\in M\right\} \\
  & & \cup \left\{[x,B],[y,A]),([y,B],[x,A]) \mid (x,y)\in E\setminus M\right\} \\
  & & \cup \left\{(s,[v,B]), ([v,A],t) \mid v\in V_M\right\}.
\end{eqnarray*}

\noindent
Analogously to the bipartite case, we have directed the edges in $M$
``from $A$ to $B$'' and the edges in $E\setminus M$ ``from $B$ to $A$''.
Since the distinct nodes $[v,A]$ and $[v,B]$ in $V'$ correspond to the same
node $v$ in $V$, it does not suffice to construct a simple path from $s$ to $t$
in $G_M$ for finding an $M$-augmenting path in $G$. Hence, we define strongly
simple paths in $G_M$ which cannot contain both nodes $[v,A]$ and $[v,B]$, for
all $v \in V$. A path $P$ in $G_M$ is {\em strongly simple} if
\begin{enumerate}
\item[a)] 
$P$ is simple, and 
\item[b)] 
$\forall [v,A] \in V': [v,A] \in P \Rightarrow [v,B]\not\in P$.
\end{enumerate}
Now we can formulate the reachability problem in $G_M$ which is equivalent to 
the problem of finding an $M$-augmenting path in $G$.

\begin{theo} \label{theo1}
Let $G=(V,E)$ be an undirected graph, $M\subseteq E$ be a matching, and
$G_M = (V',E_M)$ be defined as above. Then there exists an $M$-augmenting path
in $G$ if and only if there exists a
strongly simple path from $s$ to $t$ in $G_M$.
\end{theo}
{\bf Proof:}
``$\Rightarrow$'': Let $Q=w_1,w_2,\ldots,w_{l-1},w_l$ be an
$M$-augmenting path in $G$. Then $w_i\not= w_j$, $1\leq i<j\leq l$,
and $w_1,w_l\in V_M$. Hence, by the construction of $G_M$,
$ Q' = s,[w_1,B],[w_2,A],\ldots,[w_{l-1},B],[w_l,A],t$
is a strongly simple path in $G_M$.

\medskip
\noindent 
``$\Leftarrow$'':
Let $P=s,[v_1,B],[v_2,A],[v_3,B],\ldots,
[v_{k-1},B],[v_k,A],t$ be a strongly simple path in $G_M$. Then
$v_i\not= v_j$, $1\leq i<j\leq k$, and $v_1,v_k\in V_M$. Hence,
$P'=v_1,v_2,\ldots,v_k$ is an $M$-augmenting path in $G$.
\QED

\medskip
Because of Theorem \ref{theo1}, the reachability problem equivalent to the problem
of finding an $M$-augmenting path in $G$ is to find a 
strongly simple path from $s$ to $t$ in $G_M$.

\subsection{The solution of the reachability problem}

Depth-first search (DFS) finds simple paths in a directed graph. Hence, we 
cannot
use DFS directly for the solution of the reachability problem in $G_M$. We will
modify the usual DFS such that the modified depth-first search (MDFS) finds
precisely the strongly simple paths in $G_M$.
Let $[v,\overline{A}] := [v,B]$ and $[v,\overline{B}] := [v,A]$. Remember that
a DFS partitions the edges of the graph into four categories \cite{AHU}.
Similarly, the edges of $G_M$ are partitioned into five categories by an
MDFS of $G_M$:
\begin{enumerate}
\item {\em Tree edges\/}, which are edges leading to new nodes $[v,X]$,
$X\in \{A,B\}$ for which
$[v,\overline{X}]$ is not a predecessor during the search.
\item {\em Weak back edges\/}, which are edges leading to new nodes $[v,A]$ for
which $[v,B]$ is a predecessor during the search.
\item {\em Back edges\/}, which go from descendants to ancestors during the 
search.
\item {\em Forward edges\/}, which go from ancestors to proper descendants but
are not tree edges.
\item {\em Cross edges\/}, which go between nodes that are neither ancestors
nor descendants of one another during the search.
\end{enumerate}
Like DFS, MDFS uses a stack $K$ for the organization of the search. Analogously
to DFS, the MDFS-stack $K$ defines a tree, the {\em MDFS-tree} $T$. Before
describing MDFS in detail, we shall describe the algorithm informally. TOP($K$)
denotes the last node added to the MDFS-stack $K$. In each step, MDFS considers
an edge $({\rm TOP}(K), [w,Y])$ which has not been considered previously.
Let $e=([v,X],[w,\overline{X}])$ be the edge under consideration. We 
distinguish two cases.

\begin{itemize}
\item[1.] $X=A$, i.e.\ $(v,w)\in M$.\hfill {\em tree edge}
\item[2.] $X=B$, i.e.\ $(v,w)\in E\setminus M$
      \vspace{-0.2cm}  
      \begin{itemize}
        \item[2.1] $[w,A]\in K$ \hfill {\em back edge}
        \item[2.2] $[w,A]\not\in K$ but $[w,B]\in K$
                \begin{itemize}
                \item[i)] $[w,A]$ has been in $K$ previously \hfill {\em cross edge}
                \item[ii)] $[w,A]$ has not been in $K$ previously \hfill {\em weak back edge}
                \end{itemize} 
        \item[2.3] $[w,A] \not\in K$ and $[w,B]\not\in K$
                \begin{itemize}
                \item[i)] $[w,A]$ has been in $K$ previously \hfill {\em
forward} or {\em cross edge}
                \item[ii)] $[w,A]$ has not been in $K$ previously \hfill {\em tree edge}
                \end{itemize}
        \end{itemize}
\end{itemize}
MDFS differs from DFS only in Cases 2.2.ii and 2.3.i. Next, we shall discuss
both of these cases.

\medskip
\noindent
{\bf Case 2.2.ii:} Since $[w,A]$ has not been in $K$ previously, DFS would 
perform the operation PUSH($[w,A]$). Since $[w,B]\in K$ and
MDFS should only construct strongly simple paths in $G_M$, the operation PUSH($[w,A]$) is 
not performed by MDFS.

Note that the path $P = s,P_1,[w,B],P_2,[v,B]$ defined by the MDFS-stack $K$ is strongly
simple. Hence, the path $Q = [x,A], P_{22},[v,B],[w,A]$ where $P_2 = P_{21},[x,A],P_{22}$ 
is strongly simple for each node $[x,A]$ on $P_2$. We say then that MDFS has {\em found\/}
the strongly simple path $Q$ from $[x,A]$ to $[w,A]$.
Since the path $Q$ is above the node $[w,B]$ with respect to the MDFS-stack $K$, 
after the execution of the operation $POP([w,B])$,
no node on $Q$ is in the MDFS-stack $K$. Moreover, as we shall prove later, for all nodes 
$[z,X]$ on $P_2$ the operations $PUSH([z,X])$, $POP([z,X])$, $PUSH([z,\overline{X}])$
and $POP([z,\overline{X}])$ have been performed before the operation $POP([w,B])$.

\medskip
\noindent
{\bf Case 2.3.i:} Since $[w,A]$ has been in $K$ before, DFS would perform no 
PUSH-operation. But the different treatment of Case 2.2.ii can cause the
following situation: 
MDFS has found a strongly simple path $Q = [w,A],Q',[u,A]$ from the node $[w,A]$
to a node $[u,A]$ but at that moment, the node $[u,B]$ was below $[w,A]$ in the MDFS-stack
$K$ such that the operation PUSH($[u,A]$) has not been performed. But now, 
$[u,B]\not\in K$.

As we shall prove later, the paths $P$ from $s$ to $[v,B]$ and $Q$ from
$[w,A]$ to $[u,A]$ are {\em strongly disjoint\/}; i.e.\ there is no node $[r,X]$
on $P$, $X \in \{A,B\}$ such that $\{[r,A],[r,B]\} \cap Q \not= \emptyset$. 
Since MDFS has found a strongly simple path $P,Q$ from $s$ to $[u,A]$, MDFS now performs 
the operation PUSH($[u,A]$).

\medskip
Note that with respect to depth-first search, the DFS-stack contains exactly
the current search path. With respect to the modified depth-first search,
the situation is different. In Case 2.3.i, the node $[u,A]$ is pushed. But
to obtain a current search path, between the nodes $[v,B]$ and $[u,A]$,
we have to insert any path $[w,A],Q',[u,A]$ which has been found by MDFS.
Since we do not want to forget the information about the first node $[w,A]$
on the path which we add between the nodes $[v,B]$ and $[u,A]$, we 
create the artificial tree edge $([v,B],[u,A])_{[w,A]}$. Such an edge is
called {\em extensible edge}. It is possible that there exists various such
paths $Q'$. Hence, after the execution of PUSH$([u,A])$, the number of
corresponding current search paths can increase. 
Whenever we say that we consider {\em a current search path\/} we mean that we can
take an arbitrary corresponding current search path. If we add to the
constructed MDFS-tree $T$ all forward, back, cross and weak back edges and  
every extensible edge $([v,B],[u,A])_{[w,A]}$ is replaced by all strongly simple 
paths $Q = [v,B],[w,A],Q',[u,A]$, then
we obtain the {\em expanded current MDFS-tree\/} $T_{exp}$.

We say that MDFS has {\em constructed\/} a path $P$ if the MDFS-stack $K$ contains the 
path $P$ where each extensible edge $([v,B],[u,A])_{[w,A]}$ in $K$ is replaced by one
of the strongly simple paths $Q = [v,B],[w,A],Q',[u,A]$ which replace this extensible
edge in $T_{exp}$. 
We say that MDFS has {\em formed\/} a strongly simple path $P$ if $T_{exp}$ contains 
$P$. We say that MDFS has {\em found\/} a strongly simple path $P',[v,B],[w,A]$ if 
the path $P',[v,B]$ is formed by MDFS and the edge $([v,B],[w,A])$ is a considered 
weak back edge.
Next we shall describe MDFS in detail. 
We have to solve the following
problem: How to find the node $[u,A]$ in Case 2.3.i?
For the solution of this problem, we assume that MDFS is organized such that 
for all nodes $[w,A]\in V'$, the following holds true:

\medskip
After performing the operation POP($[w,A]$), MDFS has always computed a
set $L_{[w,A]}$ of nodes such that $L_{[w,A]}$ contains exactly those
nodes $[u,A]\in V'$ satisfying the requirements that  \label{WM1}
\begin{enumerate}
\item MDFS has found a path $P = [w,A],Q,[u,A]$ with $[u,B] \not\in Q$,
\item PUSH($[u,A]$) has never been performed, and
\item POP($[u,B]$) has been performed.
\end{enumerate}
Before the execution of POP$([w,A])$, we fix $L_{[w,A]} := \emptyset$. 

\medskip
In the description of MDFS we assume for all $[w,A]\in V'$ that
$L_{[w,A]}$ is computed correctly. As we shall prove later, it will always hold
that $|L_{[w,A]}| \leq 1$.
The computation of $L_{[w,A]}$ as well
as an efficient implementation of MDFS, can be found in Section~\ref{MDFSim}.
For $[v,X] \in V'$, $N[v,X]$ denotes the adjacency list of $[v,X]$.
Note that after the POP of the head $[w,B]$ of a matched edge $([v,A],[w,B])$,
the POP of the tail $[v,A]$ of this edge has also to be performed. Hence, at the
end of the procedure SEARCH, we shall have two POP-operations.

\medskip
\noindent
{\bf Algorithm 2 (MDFS)} \\
{\bf Input:} $G_M=(V',E_M)$. \\
{\bf Output:} A strongly simple path $P$ from $s$ to $t$, if such a path exists.\\
{\bf Method:}
\vspace{-0.3cm}
\begin{tabbing}
AA \= AA \= \= \kill
Initialize the stack $K$ to be empty; \\
PUSH($s$); \\
{\bf while} $K\not=\emptyset$ {\bf and} no path from $s$ to $t$ is constructed \\
\> {\bf do} \\
\> \> SEARCH \\
\> {\bf od}.
\end{tabbing}
\noindent
SEARCH is a call of the following procedure. 

\begin{tabbing}
(Case 2.3 ii)\= AA \= AA \= AA \= AA \= AA \= AA \= AA \= AA \= \kill 
\> {\bf procedure} SEARCH; \\
\> {\bf if} TOP$(K)$ = $t$ \\
\> {\bf then} \\
\> \> reconstruct a strongly simple path $P$ from $s$ to $t$ \\ 
\> \> which has been constructed by the algorithm \\
\> {\bf else} \\ 
\> \> mark TOP$(K)$ ``pushed''; \\
\> \> {\bf for} all nodes $[w,Y]\in N[TOP(K)]$ \\
\> \> {\bf do} \\
\> \> \> {\bf if} $Y$ = $B$ \\
(Case 1) \> \> \> {\bf then} \\
\> \> \> \> PUSH($[w,B]$); \\
\> \> \> \> SEARCH \\
(Case 2) \> \> \> {\bf else} \\
\> \> \> \> {\bf if} $[w,A]\in K$ \\
(Case 2.1) \> \> \> \> {\bf then} \\
\> \> \> \> \> no PUSH-operation is performed \\
\> \> \> \> {\bf else} \\
\> \> \> \> \>{\bf if} $[w,B]\in K$ \\
(Case 2.2) \> \> \> \> \> {\bf then} \\
\> \> \> \> \> \>no PUSH-operation is performed \\
(Case 2.3) \> \> \> \> \> {\bf else} \\
\> \> \> \> \> \> {\bf if} $[w,A]$ is marked ``pushed''  \\
(Case 2.3.i) \> \> \> \> \> \> {\bf then} \\
\> \> \> \> \> \> \>{\bf while} $L_{[w,A]} \not= \emptyset$  \\
\> \> \> \> \> \> \> {\bf do} \\
\> \> \> \> \> \> \> \> choose any $[u,A]\in L_{[w,A]}$; \\
\> \> \> \> \> \> \> \> PUSH($[u,A]$); \\
\> \> \> \> \> \> \> \> SEARCH \\
\> \> \> \> \> \> \> {\bf od} \\
(Case 2.3.ii)\> \> \> \> \> \> {\bf else} \\
\> \> \> \> \> \> \> PUSH($[w,A]$); \\
\> \> \> \> \> \> \> SEARCH \\
\> \> \> \> \> \> {\bf fi} \\
\> \> \> \> \> {\bf fi} \\
\> \> \> \> {\bf fi} \\
\> \> \> {\bf fi} \\
\> \> {\bf od};\\
\> \> POP; \\
\> \> POP \\
\> {\bf fi}.
\end{tabbing}

\subsection{The correctness proof of MDFS}

The correctness proof of MDFS is inspired by the correctness proof of DFS.
But in contrast to DFS, the proof is difficult. The difficulties come from the 
fact that the MDFS-stack does not contain the whole current search path and the 
decisions taken by the algorithm only depend on the content of the current stack.
Hence, the proof that the algorithm constructs only strongly simple paths is
involved. First we shall prove some lemmas. The first lemma implies that the first
PUSH-operation which destroys the property ``strongly simple'' must push 
a node with second component $A$.
\begin{lemma}
As long as MDFS constructs only strongly simple paths, the following holds true:
After the operation PUSH$([v,A])$ where $v$ is not $M$-free, the
operation PUSH$([w,B])$ where $([v,A],[w,B]) \in E_M$ always follows without 
destroying the property ``strongly simple''.
\end{lemma}
{\bf Proof:} 
After the execution of the operation PUSH$([v,A])$, always the unique edge 
$([v,A],[w,B]) \in E_M$ is considered and the operation PUSH$([w,B])$ is
performed. If this operation would destroy the property ``strongly simple'',
then $[w,A]$ and hence, $[v,B]$ would be on a current search path. But then,
already the operation PUSH$([v,A])$ would have destroyed the property ``strongly
simple'', a contradiction. 
\QED

\medskip
The next lemma shows that in a certain situation MDFS constructs a path from $s$ to a 
node $[x,A]$.
\begin{lemma} \label{Lemma1}
Let $[u,B]\in V'$ be a node for which MDFS performs the operation
PUSH$([u,B])$. Furthermore, at the moment when POP$([u,B])$ is performed, only strongly 
simple paths have been constructed by MDFS.
Let $[x,A]\in V'$ be a node such that at the moment when PUSH$([u,B])$ is 
performed, there is a strongly simple path
$P\space =\space [u,B],[v,A],Q,[x,A]$ with $[z,X],[z,\overline{X}] \not\in K$
for all $[z,X] \in P$. Then PUSH$([x,A])$ has been performed before POP$([u,B])$.
\end{lemma}
{\bf Remark:} Lemma~\ref{Lemma1} implies that either PUSH($[x,A]$) 
and POP($[x,A]$) have been performed before the execution of PUSH($[u,B]$), 
or both operations have been performed between the operations PUSH($[u,B]$) 
and POP($[u,B]$).

\medskip
\noindent 
{\bf Proof:}
Assume that the assertion does not hold. We consider a path
$P = [u,B],[v,A],[v',B],Q',[x,A]$ of shortest length such that 
PUSH$([x,A])$ has {\em not\/} been performed before POP$([u,B])$.
It is clear that the edge $e = ([u,B],[v,A])$ has been considered before the 
execution of POP$([u,B])$. By assumption, $[v,A] \not\in K$ and $[v,B] \not\in K$
at the moment when $e$ is considered. Hence, MDFS is in Case 2.3.

If the operation PUSH$([v,A])$ is performed according to this consideration of 
edge $e$, then PUSH$([v',B])$ would be the next operation performed by the
algorithm. By assumption, $P$ is a shortest path such that the assertion
is not fulfilled. Therefore, PUSH$([x,A])$ has been performed before POP$([v',B])$, 
and hence, before POP$([u,B])$, a contradiction. 

If the operation PUSH$([v,A])$ is not performed 
according to this consideration of edge $e$, then MDFS is in Case 2.3.i and performs 
the corresponding while-statement. Consider the moment when MDFS terminates this 
while-statement; i.e., $L_{[v,A]} = \emptyset$.
Let $[z,A]$ be the first node on $P$ for which PUSH$([z,A])$ has not been
performed. Since $[x,A]$ has this property, the node $[z,A]$ exists.
Let $P = [u,B],[v,A],Q_1,[y,B],[z,A],Q_2,[x,A]$. By construction, each node on
$P_1 = [u,B],[v,A],Q_1,[y,B]$ is pushed and each edge on $P_1$ is considered.
Hence, all these nodes and edges are in $T_{exp}$ such that $P_1$ is formed by MDFS.
Furthermore, the edge $([y,B],[z,A])$ is a considered weak back edge. Therefore,
MDFS has found the path $P_1,[z,A]$. By assumption, PUSH$([z,A])$ has never been 
performed. Since $[z,B] \not\in K$ when PUSH$([u,B])$ is performed it holds that
POP$([z,B])$ has been performed.
Hence, $[z,A] \in L_{[v,A]}$, and hence, $L_{[v,A]} \not= \emptyset$. But this 
contradicts $L_{[v,A]} = \emptyset$ such that the lemma is proved.
\QED

\medskip
The following lemma shows that in a certain situation some nodes are 
already pushed.
For the proof of the lemma, we need the notation of the so-called 
back-path $r(S)$ of a path $S$ in $G_M$. Essentially, the back-path of $S$ is
obtained by changing the direction of the edges on $S$ and running $S$ backwards. 
To get a formal definition for $r(S)$, we denote for $w \in V'$
$$ r(w) :=
\left\{\begin{array}{cl}
[v,\overline{X}] & {\rm if\ } w=[v,X] \\
t                & {\rm if\ } w=s \\
s                & {\rm if\ } w=t.
\end{array}\right.
$$
Let $S = w_1,w_2,\ldots,w_k$ be a path in $G_M$. The {\em back-path\/}
$r(S)$ of $S$ is defined by
$$ r(S) := r(w_k), r(w_{k-1}), \ldots,r(w_1).$$
The following lemma is a direct consequence of Lemma \ref{Lemma1}.
\begin{lemma} \label{Lemma3}
Let $[u,B]\in V'$ be a node for which MDFS performs the operation
PUSH$([u,B])$. 
Furthermore, at the moment when POP$([u,B])$ is performed, only strongly simple 
paths have been constructed by MDFS. If there exists a strongly simple path
$P=[v,A],Q,[w,B]$ such that at the moment when PUSH$([u,B])$ is performed,  
$[z,X],[z,\overline{X}] \not\in K$, for all $[z,X] \in P$, and 
$([u,B],[v,A]), ([w,B],[u,A]) \in E_M$ then for all $[z,X] \in P$, the 
operations PUSH$([z,X])$ and PUSH$([z,\overline{X}])$ have been performed 
before the execution of the operation POP$([u,B])$.
\end{lemma}
{\bf Proof:} 
For $[z,X] \in P$, the assertion follows by an application of Lemma \ref{Lemma1}
to the path $[u,B],P$. With respect to $[z,\overline{X}]$ where $[z,X] \in P$, 
we apply Lemma \ref{Lemma1} to the path $[u,B],r(P)$.
\QED

\medskip
Note that by the definition of $L_{[u,A]}$, $|L_{[u,A]}| > 0$ implies that PUSH$([u,A])$ 
and POP$([u,A])$ have been performed. The following lemma extracts properties of
the algorithm MDFS which enable us to prove the correctness and to develop an efficient 
implementation of MDFS.

\begin{lemma} \label{Lemma4}
MDFS maintains the following invariants:
\vspace{-0.2cm}
\begin{enumerate}
\item 
MDFS constructs only strongly simple paths.
\item 
$|L_{[w,A]}| \leq 1$, for all $[w,A] \in V'$.
\item
Assume that the algorithm performs the assignment $L_{[w,A]} := [u,A]$. 
Then after the execution of PUSH$([u,A])$ always $L_{[w,A]} = L_{[u,A]}$. 
\end{enumerate}
\end{lemma}
{\bf Remark:} Invariant 2 and Invariant 3 are not needed for the correctness
proof. But we shall need these invariants for an efficient 
implementation of the algorithm. Moreover, the proof of Invariant 1 is easier 
if we prove all invariants simultaneously.

\medskip
\noindent
{\bf Proof:} Consider the first situation in which one of the three invariants
is not maintained. Three cases have to be considered.

\medskip
\noindent
{\bf Case 1:} Invariant 1 is not maintained.

\medskip
Only a PUSH-operation can destroy the property ``strongly simple''. Note that
a PUSH-operation cannot affect Invariant 2 or Invariant 3.
Lemma 1 implies that this PUSH-operation occurs during the consideration
of an edge $e = ([v,B],[w,A])$. 

If $[w,A]$ is not marked ``pushed'', then Case 2.3.ii applies and
PUSH$([w,A])$ is performed. Since $[w,B] \not\in K$, 
the only possible situation in which this PUSH-operation destroys the
property ``strongly simple'' is the following:
On a current search path there is a subpath $Q$ which is caused by an 
application of Case 2.3.i of MDFS such that $[w,B] \in Q$.
Hence, there exists $[u,A] \in V'$ such that the addition of $Q$ to this 
current search path is caused by the operation PUSH$([u,A])$. By construction, 
the assumptions of Lemma \ref{Lemma3} are fulfilled with respect to $[u,B]$ and $[w,B]$ on $P$. 
Hence, by Lemma \ref{Lemma3}, PUSH$([w,A])$ has been performed {\em before\/} 
POP$([u,B])$, and hence, {\em before\/} PUSH$([u,A])$, a contradiction. 

Hence, $[w,A]$ is marked ``pushed'' such that Case 2.3.i of MDFS 
applies. By Invariant 2, $|L_{[w,A]}| \leq 1$. We thus write 
$L_{[w,A]} = [u,A]$ instead of $L_{[w,A]} = \{[u,A]\}$. 
By Case 2.3.i, for the node $[u,A] = L_{[w,A]}$, the operation PUSH$([u,A])$
is performed such that MDFS extends the current search paths by a path $[w,A],Q,[u,A]$, 
but only $[u,A]$ is pushed. 
By the definition of $L_{[w,A]}$ and by Lemma 3, the operations PUSH$([z,X])$,
POP$([z,X])$, PUSH$([z,\overline{X}])$, and POP$([z,\overline{X}])$ have been 
performed, for all $[z,X] \in Q$ such that none of these nodes is contained
in the current MDFS-stack $K$. Hence, the only possible situation in which
PUSH$([u,A])$ destroys the property ``strongly simple'' is the following:
There is a node $[p,X] \in [w,A],Q,[u,A]$, and a subpath $Q'$ of a current
search path which is caused by an application of Case 2.3.i such that
$[p,X] \in Q'$ or $[p,\overline{X}] \in Q'$. Since one end node of an edge in 
the current matching uniquely determines the other end node, we can choose 
$[p,X]$ such that $[p,A] \in Q'$.

Consider the node $[u',A] \in K$ such that PUSH$([u',A])$ is the operation which adds 
the subpath $Q'$ to this current search path. Therefore, immediately before the execution
of PUSH$([u',A])$, $L_{[p,A]} = [u',A]$. Hence, by Invariant 3,
after the execution of PUSH$([u',A])$, always $L_{[p,A]} = L_{[u',A]}$. By
the choice of $[p,A]$, $L_{[p,A]} = [u,A]$, and hence, $L_{[u',A]} = [u,A]$ in 
the situation under consideration. Hence, 
POP$([u',A])$ has been performed such that $[u',A] \not\in K$, a contradiction. 

\medskip
\noindent
{\bf Case 2:} Invariant 2 is not maintained.

\medskip
Then there exists $[w,A],[p_1,A],[p_2,A] \in V'$ such that 
$L_{[w,A]} = \{[p_1,A]\}$ before the execution of POP$([p_2,B])$ and 
$L_{[w,A]} = \{[p_1,A],[p_2,A]\}$  after the execution of POP$([p_2,B])$.
Hence, MDFS has found a path $P_1 = [p_1,B],Q,[p_1,A]$ with $[w,A] \in Q$ and 
found a path $P_2 = [p_2,B],Q',[p_2,A]$ with $[w,A] \in Q'$.

If MDFS has found the path $P_2$ after the execution of POP$([p_1,B])$, then
$[w,A]$ can only be added to $Q'$ in the following way:
An operation PUSH$([u,A])$, caused by an application of Case 2.3.i
with respect to a node $[v,A]$ (i.e., $[u,A] \in L_{[v,A]}$) is performed 
such that the current search path is extended by a path $[v,A],\tilde{Q},[u,A]$
with $[w,A] \in \tilde{Q}$.
But then, $[u,A] \in L_{[w,A]}$ before the execution of PUSH$([u,A])$. 
PUSH$([u,A])$ is performed after POP$([p_1,B])$. Hence, $[u,A],[p_1,A]\in 
L_{[w,A]}$ between the execution of these two operations.
This contradicts the assumption that we consider the situation in
which Invariant 2 is not maintained {\em for the first time\/}.

Hence, MDFS has found the path $P_2$ before the execution of POP$([p_1,B])$. 
Note that $[p_1,B] \not\in Q'$. Otherwise, by Lemma 3, 
PUSH$([p_1,A])$ is performed before POP$([p_2,B])$, and hence, $[p_1,A] 
\not\in L_{[w,A]}$ after POP$([p_2,B])$. 
Let $[r,A]$ be the first node on $Q'$ such that $[r,A] \in Q$ or $[r,B] \in
Q$. Since node $[w,A]$ has this property, the node $[r,A]$ exists. Let 
$$Q' = Q'_1,[r,A],Q'_2 \quad \mbox{ and } \quad
Q = \left\{ \begin{array}{ll}
                Q_1,[r,A],Q_2 & \mbox{if $[r,A] \in Q$} \\
                Q_1,[r,B],Q_2 & \mbox{if $[r,B] \in Q$.}
              \end{array}
      \right.$$   
Consider the path
$$R = \left\{ \begin{array}{ll}
                Q'_1,[r,A],Q_2,[p_1,A] & \mbox{if $[r,A] \in Q$} \\
                Q'_1,[r,A],r(Q_1),[p_1,A] & \mbox{if $[r,B] \in Q$.}
              \end{array}
      \right.$$
Then Lemma 2 applies with respect to $[p_2,B]$, $[p_1,A]$ and the
strongly simple path $R$. Hence, PUSH$([p_1,A])$ is performed
before POP$([p_2,B])$, and hence, $[p_1,A] \not\in L_{[w,A]}$
after POP$([p_2,B])$, a contradiction.

\medskip
\noindent 
{\bf Case 3:} Invariant 3 is not maintained.

\medskip
After the execution of PUSH$([u,A])$, it holds that
$L_{[w,A]} = L_{[u,A]} = \emptyset$. 
We shall prove that $L_{[w,A]} = L_{[u,A]}$ after the next 
POP-operation which changes $L_{[w,A]}$ or $L_{[u,A]}$. Then, the assertion 
follows because of Invariant 2 and the transitivity of the relation $=$.
Let POP$([p,B])$ be the next POP-operation which enlarges $L_{[w,A]}$ or
$L_{[u,A]}$. $K_{[w,A]}$ denotes the current MDFS-stack, directly after the 
execution of PUSH$([w,A])$. Let $K' = K_{[w,A]} \cap K_{[u,A]}$.
Note that $[u,B] \in K_{[w,A]} \setminus K'$.
According to the location of $[p,B]$ with respect to $K_{[w,A]}$ and to
$K_{[u,A]}$, we have to discuss three cases.

By construction, $[p,B] \not\in K_{[w,A]} \setminus K'$. Otherwise, 
POP$([p,B])$ would be performed before PUSH$([u,A])$.

Assume that $[p,B] \in K_{[u,A]} \setminus K'$. Let $[q,B]$ be the first node 
in $K_{[w,A]} \setminus K'$ such that 
$[q,A] \in K_{[u,A]} \setminus K_{[p,B]}$. Node $[q,B]$ exists since $[u,B]$
has the property that $[u,B] \in K_{[w,A]} \setminus K'$.
Consider the back-path of the path from node $[p,B]$ to node $[q,A]$. This 
back-path implies that $[q,B]$ and $[p,A]$ fulfill the assumptions of
Lemma 2. Hence, PUSH$([p,A])$ occurs before POP$([q,B])$. Since $[q,B] \in
K_{[w,A]} \setminus K'$, the operation PUSH$([p,A])$ is also performed before
POP$([p,B])$. Hence, POP$([p,B])$ can enlarge neither $L_{[w,A]}$ nor
$L_{[u,A]}$. 

It remains to consider $[p,B] \in K'$. Let $[q,B] \in K'$ be the node 
nearest to the top of $K'$ for which PUSH$([q,A])$ has not been performed at 
the moment when MDFS performs PUSH$([u,A])$. Since $[p,B]$ has this property,
$[q,B]$ exists.
By consideration of the back-path of the path from $[q,B]$ to $[u,B]$, 
it is easy to prove that MDFS finds a path from $[u,A]$ to $[q,A]$
not containing $[q,B]$. Hence, $L_{[u,A]} = [q,A]$ after the 
execution of POP$([q,B])$, and hence, 
$[q,B] = [p,B]$. Since MDFS has found a path from $[w,A]$ to $[u,A]$ which 
does not contain $[q,B]$, it holds that $L_{[w,A]} = [q,A] = [p,A]$.  
\QED

\medskip
Now, the correctness of the algorithm MDFS can easily be derived from Lemma \ref{Lemma1}
and Lemma \ref{Lemma4}.
\begin{theo}
MDFS constructs a strongly simple path from $s$ to $t$ iff such a path exists.
\end{theo}
{\bf Proof:}
Assume $P=s,[v'_0,B],[v_1,A],[v'_1,B],\ldots,[v'_{r-1},B],[v_r,A],t$ is a strongly simple path 
from $s$ to $t$. It is clear that MDFS considers the
edge $(s,[v'_0,B])$ and performs the operation PUSH($[v'_0,B]$). (Note that 
$v'_0$ is $M$-free.)
Hence, $[v'_0,B]$, $[v_r,A]$ fulfill the assumptions of Lemma~\ref{Lemma1}
with respect to the path $[v'_0,B],[v_1,A],\ldots,[v'_{r-1},B],\linebreak[0]
[v_r,A]$.
By Lemma~\ref{Lemma1}, MDFS performs PUSH($[v_r,A]$) and hence, 
PUSH($t$). Therefore, MDFS constructs a path from $s$ to $t$.
By Invariant 1 of Lemma \ref{Lemma4}, MDFS constructs only strongly simple paths.
\QED

\subsection{An implementation of MDFS} \label{MDFSim}

Now we shall describe how to get an efficient implementation of MDFS. Only 
two parts of the algorithm are nontrivial to implement.
\begin{enumerate}
\item The manipulation of $L_{[w,A]},\; [w,A]\in V'$.
\item The reconstruction of a strongly simple path~$P$ from $s$ to $t$ which
 is constructed by the algorithm.
\end{enumerate}
For the solution of both subproblems it is useful to perform the
POP-operations not explicitly and to maintain the whole MDFS-tree $T$.
This can be done as follows:
The data structure is a tree $T$. A pointer $TOP$ always points to TOP$(K)$ 
in $T$. The current MDFS-stack $K$ is represented by the unique path from the 
root $s$ of $T$ to TOP$(K)$ in $T$. For performing the operation POP, the
pointer $TOP$ is changed such that it points to the unique direct predecessor 
in $T$. When we perform a PUSH-operation, the node in $T$ to which TOP$(K)$ 
points obtains a new leaf. After the PUSH-operation TOP$(K)$ points to this
new leaf.

Invariant 2 and Invariant 3 are the key for the efficient implementation of 
MDFS. Now we shall describe the update of $L_{[w,A]}$. By the definition of
$L_{[w,A]}$, we only have to change $L_{[w,A]}$ after a PUSH- or after
a POP-operation. More exactly, we have to perform after PUSH($[u,A]$) the 
operation $L_{[w,A]} := \emptyset$ if $L_{[w,A]} = [u,A]$ and after POP($[u,B]$)
the operation $L_{[w,A]} := [u,A]$ if PUSH($[u,A]$) has never been performed and 
MDFS has found a path $P = [w,A],Q,[u,A]$ for which $[u,B] \not\in Q$.

After the execution of POP$([u,B])$, if PUSH$([u,A])$ has never been performed, 
MDFS needs all nodes $[w,A]$ for which a path $P = [w,A],Q,[u,A]$ such that 
$[u,B] \not\in Q$ has been found by MDFS.
This can easily be done by any graph search method like depth-first search on the
current $T_{exp}$, starting at node $[u,A]$ and running the considered edges backwards. 
When the node $[u,B]$ is reached, a backtrack is performed. But with respect
to efficiency, it is useful to investigate the properties of MDFS and to
refine the backward graph search. 

First, we shall characterize the paths $P = [w,A],Q,[u,A]$ with $[u,B] \not\in
Q$, found by MDFS. Let $P = e_1,e_2, \ldots ,e_t$. Then, the following 
properties are fulfilled:
\begin{enumerate}
\item 
$e_t$ is a weak back edge.
\item
If we start in edge $e_t$ and consider $P$ backwards, then we see a nonempty sequence
of tree edges followed by a single cross, forward or back edge, followed by a 
nonempty sequence of tree edges followed by a single cross, forward or 
back edge, and so on.
\end{enumerate}
Hence, after the execution of POP$([u,B])$, we need the following sets
of nodes:
$$R_{[u,A]} := \left\{[v,B]\in V'\mid ([v,B],[u,A])
  \mbox{ is a weak back edge}\right\}$$
and for some $[q,A] \in V'$
$$
E_{[q,A]} := \{[v,B] \in V' \mid ([v,B],[q,A]) \mbox{ is a cross, forward,
              or back edge}\}.
$$
According to Invariant 3, during the backward search, some subpaths can be
skipped over. Therefore, we need the following set of nodes
$$\label{WM2}
D_{[q,A]} := \{[p,A] \in V' \mid L_{[p,A]} = [q,A] \mbox{ previously}\}.
$$
By Invariant 3, $D_{[q,A]} \subseteq D_{[q',A]}$ implies $L_{[q,A]} =
L_{[q',A]}$. 
We say that $D_{[q,A]}$ is {\em current\/} if $D_{[q,A]} \not\subseteq
D_{[q',A]}$, for all $[q',A] \in V'\setminus \{[q,A]\}$.
According to Invariant 3, we can compute $L_{[p,A]}$ in the following way.
\begin{enumerate}
\item Compute $[q,A]$ such that $[p,A] \in D_{[q,A]}$, and $D_{[q,A]}$ is 
current.
\item If $[q,A]$ does not exist, then $L_{[p,A]} = \emptyset$. Otherwise, 
$$L_{[p,A]} = \left\{ \begin{array}{ll}
                        [q,A] & \mbox{if PUSH$([q,A])$ has never been performed} \\
                        \emptyset & \mbox{otherwise.}
                      \end{array}
              \right.$$
\end{enumerate} 
As described above, a correct manipulation of the current sets $D_{[q,A]}$ 
allows the solution of the first subproblem. Note that by Invariant 2
of Lemma \ref{Lemma4}, each 
$[p,A] \in V'$ is contained in at most one current set $D_{[q,A]}$. 
If during the backward search a node $[p,A]$ is met for which $L_{[p,A]} \not= \emptyset$
previously, some nodes can be skipped over. Hence, we have also to know 
if $L_{[p,A]} \not= \emptyset$ previously. This will be realized by the correct 
update of the following set
$$L := \{[p,A] \in V' \mid L_{[p,A]} \not= \emptyset \mbox{ previously}\}.$$
Now we can give a detailed description of the backward search which will be 
performed after POP$([u,B])$.
The consideration of those paths $P = [w,A],Q,[u,A]$ with $[u,B] \not\in Q$
is done in several rounds.
In the first round, we construct backwards all paths {\em without any\/} cross,
forward, or back edge.
In the second round, all paths with {\em exactly one} such edge are constructed
implicitly, and so on.
Let $T_v$ denote the nodes in $T_{exp}$ which have been already considered during 
the backward search. If a node in $T_v$ is considered again, the search has not 
to be continued at that node.
In the first round, we consider the weak back edges $([v,B],[u,A])$. In the $i$th
round, $i > 1$, we consider those edges $([v,B],[q,A]) \in E_{[q,A]}$ for which
$L_{[q,A]} = [u,A]$ is computed in the $(i-1)$st round.
Starting in node $[v,B]$, we follow backwards the tree edges as long as 
a node in $T_v \cup \{[u,B]\}$ is reached. If we reach a node $[p,A] \in L$, then 
we compute the current $D_{[r,A]}$ such that $[p,A] \in D_{[r,A]}$ and we jump to
$[r,B]$ for the continuation of the backward search. Since we perform a backward
search, $[r,B]$ is switched to $[r,A]$. According to 
Invariant 3, $L_{[x,A]} = L_{[r,A]}$ and hence, $L_{[x,A]} = [u,A]$ for all 
$[x,A] \in D_{[r,A]}$. 

For the organization of the backward search, we use a queue $\cal{Q}$ which contains
the start nodes of the next round. During Round $i$, the start nodes of Round $i+1$ 
are added to the end of the queue. Since the data structure is a queue, all start
nodes of a round are treated before the first start node of the next round is taken
away from $\cal{Q}$.

For the reconstruction of a strongly simple path from $s$ to $t$ constructed
by the algorithm, we have to know the non-tree edges used on the path. Hence, 
for all $[r,A]\in V'$ we use a variable $P_{[r,A]}$ to store the needed information
with respect to the node $[r,A]$. This means, we store in variable $P_{[r,A]}$ that non-tree
edge which concludes the block of tree edges which contains the tree edge with end 
node $[r,A]$ at the moment when $L_{[r,A]} \not= \emptyset$ for the first time 
during the backward search.

The implementation of MDFS must be done with attention to the correct 
manipulation of the
sets $D_{[q,A]}, R_{[q,A]}$, and $E_{[q,A]}$. The following table describes in
terms of the case of MDFS, and in terms of the operation which is 
performed, how MDFS has to update these sets.
\begin{center}
\begin{tabular}{|l|l|} \hline
{\em case, operation\/}  & {\em set updating\/} \\ \hline
Case 1  &  no update \\
Case 2.1  &  $E_{[w,A]}$ := $E_{[w,A]}\cup \{[v,B]\}$  \\
Case 2.2.i  & $E_{[w,A]}$ := $E_{[w,A]} \cup \{[v,B]\}$ \\
Case 2.2.ii  &  $R_{[w,A]}$ := $R_{[w,A]}\cup \{[v,B]\}$  \\
Case 2.3.i  &                                          \\
$L_{[w,A]} \not= \emptyset$  & no update  \\
$L_{[w,A]}=\emptyset$ & $E_{[w,A]}$ := $E_{[w,A]}\cup \{[v,B]\}$ if $[w,A] 
\not\in L$ \\
Case 2.3.ii  & no update   \\
PUSH$([u,A])$ & no update \\
POP$([v,B])$  & $D_{[v,A]}$ := $\{[p,A] \mid$ MDFS has found a path from \\
  &  $[p,A]$ to $[v,A]$ not containing $[v,B] \}$  \\
\hline
\end{tabular}
\end{center}
In Case 2.1, it is clear that $[w,A] \not\in L$ since POP$([w,A])$ has not been 
performed. In Case 2.2.i, $[w,A] \not\in L$ follows directly from $[w,B] \in
K$ and Lemma 1.
Note that in Case 2.3.i, subcase $L_{[w,A]}\not= \emptyset$, we have to 
store the information that edge $([v,B],[w,A])$ is used. In the 
implementation, we accomplish this by adding the edge $([v,B],[w,A])$ to the node 
$[v,B]$ in $K$.
Then we obtain the {\em expanded node} $\langle ([v,B],[w,A]);[v,B]\rangle$.
The considerations above lead to the following implementation of the procedure
SEARCH.

\medskip

\begin{tabbing}
(Case 2.3.ii)\= AA \= AA \= AA \= AA \= AA \= AA \= AA \= AA \= AA \= \kill
\> {\bf procedure} SEARCH; \\
\> {\bf if} TOP$(K)$ = $t$ \\
\> {\bf then} \\
\> \> reconstruct a strongly simple path $P$ from $s$ to $t$ \\
\> \> which has been constructed by the algorithm \\
\> {\bf else} \\
\> \> mark TOP$(K)$ ``pushed''; \\
\> \> {\bf for} all nodes $[w,Y] \in N[TOP(K)]$ \\
\> \> {\bf do} \\
\> \> \> {\bf if} $Y$ = $B$ \\
(Case 1) \> \> \> {\bf then} \\
\> \> \> \> PUSH$([w,B])$; \\
\> \> \> \> SEARCH \\
(Case 2) \> \> \> {\bf else} \\
\> \> \> \> {\bf if} $[w,A] \in K$ \\
(Case 2.1) \> \> \> \> {\bf then} \\
\> \> \> \> \> $E_{[w,A]}$ := $E_{[w,A]}\cup \{TOP(K)\}$ \\
\> \> \> \> {\bf else} \\
\> \> \> \> \> {\bf if} $[w,B] \in K$ \\
(Case 2.2) \> \> \> \> \> {\bf then} \\
\> \> \> \> \> \> {\bf if} $[w,A]$ is marked ``pushed'' \\
(Case 2.2.i) \> \> \> \> \> \> {\bf then} \\
\> \> \> \> \> \> \> $E_{[w,A]} := E_{[w,A]}\cup \{TOP(K)\}$ \\
(Case 2.2.ii) \> \> \> \> \> \> {\bf else} \\
\> \> \> \> \> \> \> $R_{[w,A]}$ := $R_{[w,A]} \cup \{TOP(K)\}$ \\
\> \> \> \> \> \> {\bf fi} \\
(Case 2.3) \> \> \> \> \> {\bf else} \\
\> \> \> \> \> \> {\bf if} $[w,A]$ is marked ``pushed'' \\ 
(Case 2.3.i) \> \> \> \> \> \> {\bf then}\\
\> \> \> \> \> \> \> {\bf if} $L_{[w,A]} \not= \emptyset$ \\
\> \> \> \> \> \> \> {\bf then} \\
\> \> \> \> \> \> \> \> expand TOP$(K)$ in $K$ to \\
\> \> \> \> \> \> \> \> $\langle (TOP(K),[w,A]);TOP(K)\rangle$; \\
\> \> \> \> \> \> \> \> PUSH$(L_{[w,A]})$; $L_{[w,A]} := \emptyset$;\\
\> \> \> \> \> \> \> \> SEARCH \\
\> \> \> \> \> \> \> {\bf else} \\
\> \> \> \> \> \> \> \> {\bf if} $[w,A] \not\in L$ \\
\> \> \> \> \> \> \> \> {\bf then} \\
\> \> \> \> \> \> \> \> \> $E_{[w,A]}$ := $E_{[w,A]}\cup \{TOP(K)\}$ \\
\> \> \> \> \> \> \> \> {\bf fi} \\
\> \> \> \> \> \> \> {\bf fi} \\
(Case 2.3.ii) \> \> \> \> \> \> {\bf else} \\
\> \> \> \> \> \> \> PUSH$([w,A])$; \\
\> \> \> \> \> \> \> SEARCH \\
\> \> \> \> \> \> {\bf fi} \\
\> \> \> \> \> {\bf fi} \\
\> \> \> \> {\bf fi} \\
\> \> \> {\bf fi} \\
\> \> {\bf od}; \\ 
\> \> $(\ast$ let TOP$(K)$ = $[v,B] \ast)$ \\
\> \> $L_{cur}$ := $[v,A]$; \\
\> \> $D_{L_{cur}}$ := $\emptyset$; \\
\> \> $\cal{Q}$ := $\emptyset$; \\
\> \> {\bf for} all $[q,B] \in R_{[v,A]}$ \\
\> \> {\bf do} \\
\> \> \> CONSTRL$(([q,B],[v,A]),[v,B])$; \\
\> \> {\bf od}; \\
\> \> {\bf while} $\cal{Q} \not= \emptyset$ \\
\> \> {\bf do} \\
\> \> \> remove the front node $[k,A]$ from $\cal{Q}$; \\
\> \> \> {\bf for} all $[q,B] \in E_{[k,A]}$ which have not already been \\
\> \> \> \> considered during the backward search \\
\> \> \> {\bf do} \\
\> \> \> \> CONSTRL$(([q,B],[k,A]),[v,B])$ \\
\> \> \> {\bf od} \\
\> \> {\bf od;} \\
\> POP; \\
\> POP \\
{\bf fi}. \\
\end{tabbing}
\vspace{-.3cm}
\noindent
CONSTRL is a call of the following procedure. The variable $P_{cur}$ contains 
always the non-tree edge which concludes the current block of tree edges.

\begin{tabbing}
(Case 2.3.ii)\= AA \= AA \= AA \= AA \= AA \= AA \= AA \= AA \= AA \= \kill
\> {\bf procedure} CONSTRL$(([q,B],[u,A]),[x,B])$; \\
\> $P_{cur}$ := $([q,B],[u,A])$; \\
\> $[z,B]$ := $[q,B]$; \\
\> {\bf while} no node in $T_v \cup \{[x,B]\}$ is reached \\
\> {\bf do} \\
\> \> starting in the node $[z,B]$, perform a backward search until \\ 
\> \> a node in $L \cup T_v \cup \{[x,B]\}$ is reached on the tree edges; \\
\> \> {\bf if} $[y,A] \not\in L$ is met during the backward search \\ 
\> \> {\bf then} \\
\> \> \> $D_{L_{cur}}$ := $D_{L_{cur}} \cup \{[y,A]\}$; \\
\> \> \> $L$ := $L \cup \{[y,A]\}$; \\
\> \> \> $P_{[y,A]}$ := $P_{cur}$; \\
\> \> \> add the node $[y,A]$ at the end of $\cal{Q}$ \\
\> \> {\bf fi}; \\
\> \> {\bf if} $[y,A] \in L$ is met by the backward search \\
\> \> {\bf then} \\
\> \> \> $(\ast$ Let $D_{[r,A]}$ be the current set containing $[y,A]$. $\ast)$\\
\> \> \> $D_{L_{cur}}$ := $D_{L_{cur}} \cup D_{[r,A]}$; \\
\> \> \> $[z,B]$ := $[r,B]$ \\
\> \> {\bf fi} \\
\> {\bf od}. \\
\end{tabbing}

\noindent
The reconstruction of a strongly simple path $P$ from $s$ to $t$ constructed by
the algorithm remains to be explained. 
Beginning at the end of $P$, such a path $P$ can be reconstructed by traversing
the MDFS-tree $T$ backwards. Note that $TOP$ points to the end of $P$,
and that the father of each node in $T$ is always unique.
As long as we traverse tree edges of the algorithm MDFS, we have no difficulty.
But every time when we meet a node $[u,A]$ which has been added to $P$ by an 
application of Case 2.3.i, we have to reconstruct a subpath $[w,A],Q,[u,A]$ 
which has been joined to $P$. In this situation, the considered portion of $T$
is the expanded node $\langle ([v,B],[w,A]); [v,B]\rangle$;
i.e., the structure of $T$ tells us that MDFS has applied Case 2.3.i.
It remains the reconstruction of the subpath $[w,A],Q,[u,A]$. For doing this, 
we start in the node $[w,A]$. We use $P_{[w,A]}$ to obtain the non-tree edge of MDFS, 
which finishes the block containing the tree edge with end node $[w,A]$.
Let $P_{[w,A]} = ([x,B],[y,A])$. Then $P^1_{[w,A]}$ denotes $[x,B]$, and
$P^2_{[w,A]}$ denotes $[y,A]$. First we reconstruct the block from the node
$[w,A]$ to the node $P^1_{[w,A]} = [x,B]$. Then we reconstruct the block from the node
$P^2_{[w,A]} = [y,A]$ to the node $P^1_{[y,A]}$, and so on until the node
$[u,A]$ is met. Each block can be reconstructed as the path $P$ itself.
These considerations lead to the following procedure for the reconstruction of
an augmenting path, constructed by the algorithm.
\begin{tabbing}
(Case 2.3.ii)\= AA \= AA \= AA \= AA \= AA \= AA \= AA \= AA \= AA \= \kill
\> {\bf procedure} RECONSTRPATH$(t,s)$; \\
\> $NODE_{cur}$ := $t$; \\
\> {\bf while} $NODE_{cur} \not= s$ \\
\> {\bf do} \\ 
\> \> {\bf if} $father(NODE_{cur})$ is not expanded \\
\> \> {\bf then} \\
\> \> \> $NODE_{cur}$ := $father(NODE_{cur})$ \\
\> \> {\bf else} $(\ast$ let $father(NODE_{cur}) = \langle ([v,B],[w,A]); [v,B] 
\rangle \ast)$ \\
\> \> \> RECONSTRQ$(NODE_{cur},[w,A])$; \\
\> \> \> $NODE_{cur}$ := $[v,B]$ \\
\> \> {\bf fi} \\
\> {\bf od}. \\
\end{tabbing}
\noindent
RECONSTRQ is a call of the following procedure.
\begin{tabbing}
(Case 2.3.ii)\= AA \= AA \= AA \= AA \= AA \= AA \= AA \= AA \= AA \= \kill
\> {\bf procedure} RECONSTRQ$([u,A],[w,A])$; \\
\> $ST$ := $[w,A]$; \\
\> RECONSTRPATH$(P^1_{ST},ST)$; \\ 
\> {\bf while} $P^2_{ST} \not= [u,A]$ \\
\> {\bf do} \\
\> \> $ST$ := $P^2_{ST}$; \\
\> \> RECONSTRPATH$(P^1_{ST},ST)$  \\
\> {\bf od}. \\
\end{tabbing}

\medskip
\noindent
The correctness of the manipulation of $L_{[w,A]}$, 
$[w,A]\in V'$, and the correctness of the reconstruction of the $M$-augmenting
path $P$ follow from Lemma \ref{Lemma4}, and are straightforward to prove.
The procedure RECONSTRPATH resembles standard recursive methods used for
the reconstruction of augmenting paths (see e.g. \cite{Ta}).
The time and space complexity of our implementation of MDFS remain to be
considered. It is easy to see that the time used by the algorithm MDFS is bounded
by $O(n+m)$ plus the total time needed for the manipulation of the sets
$D_{[q,A]}$, $[q,A] \in V'$.
If we use linear lists for the realization of the sets $D_{[q,A]}$ with a 
pointer to the node $[q,A]$ for each element of $D_{[q,A]}$, the 
execution time for each union operation is bounded by $O(n)$. Following the 
pointer corresponding to $[p,A]$, we can find the set containing $[p,A]$ 
in constant time.
At most $n$ union operations are performed by MDFS. 
Hence the total time used for the manipulation of the sets $D_{[q,A]}$ is 
bounded by $O(n^2)$. The time needed for the $n$ union operations can be 
reduced to $O(n \log n)$ if we use the following standard trick,
the so-called {\em weighted union heuristic\/}:

\medskip
We store with each set the number of elements of the set. A union operation 
is performed by changing the pointer of the smaller of the two sets which are
involved and updating the number of elements. Every time when the pointer with 
respect to an element is changed,
the size of the set containing this element is at least twice of the size of
its previous set. Hence, for each element, its pointer is changed at most
$\log n$-times. Hence, the total time used for all union operations is 
$O(n \log n)$. Altogether, the total time used for the augmentation of one 
augmenting path is $O(m + n \log n)$.

\medskip
If we use for the update of the sets $D_{[q,A]}$ disjoint set 
union \cite{Ta}, the total time can be bounded to be $O((m+n)\alpha(m,n))$
where $\alpha$ is the inverse Ackermann function.
Note that for each node $[p,A]$ one find operation suffices for the decision 
of $L_{[p,A]}$.
Furthermore, we can reduce these bounds to $O(m+n)$ using incremental tree set
union \cite{GaTa}.
The space complexity of MDFS is bounded by $O(m+n)$. The considerations above 
lead to the following theorem.
\begin{theo}
MDFS can be implemented such that it uses only $O(m+n)$ time and $O(m+n)$ 
space.
\end{theo}

\section{The Hopcroft-Karp approach for general graphs}

In 1973, Hopcroft and Karp \cite{HK} proved the following fact. If one computes in 
one phase a maximal set of shortest pairwise disjoint augmenting paths
and augments these paths then $O(\sqrt{n})$ such phases would suffice. 
In the bipartite case, they have described an elegant simple $O(m+n)$ implementation 
of an entire phase. Let us sketch this implementation. First they have reduced 
the problem of finding augmenting paths to a reachability problem in a directed
graph $G_M$ with two additional nodes $s$ and $t$.
Then, by performing a breadth-first search on $G_M$ with start node $s$
until the target node $t$ is reached, they have obtained a layered, directed
graph $\bar{G}_M$ for which the paths from~$s$ to~$t$ correspond exactly to
the shortest $M$-augmenting paths in $G$. Using depth-first search, they find a
maximal set of pairwise disjoint $M$-augmenting paths. Whenever an $M$-augmenting path
is found, the symmetric difference is applied to the path and the current matching, 
the path and all incident edges are deleted and the depth-first search is continued. 
Breadth-first search and depth-first search
take $O(m+n)$ time. Hence, the implementation of Hopcroft and Karp  of a phase has 
time complexity $O(m+n)$.
Since $M$-augmenting paths can be found in general graphs by a slightly modified 
depth-first search (MDFS), the following question suggests itself:
Can we get an implementation of an entire phase by performing something like
breadth-first search followed by MDFS?
We shall give an affirmative answer to this question.

\subsection{The description of a phase}

Let $G=(V,E)$ be an undirected graph, $M$ be a matching of $G$, and 
$G_M=(V',E_M)$ be the directed graph as defined in Section 2.2.
Our goal is to construct from $G_M$ a layered directed graph
$\bar{G}_M = (V',\bar{E}_M)$ such that
\begin{enumerate}
\item the $l$th layer contains exactly those nodes $[v,X]\in V'$ such that a
      shortest strongly simple path from~$s$ to~$[v,X]$ in $G_M$ has length~$l$, and
\item $\bar{G}_M$ contains all shortest strongly simple paths from~$s$ to~$t$ 
      in $G_M$.
\end{enumerate}
The {\em level} of a node $[v,X] \in V'$ is the length of a shortest strongly simple path 
from $s$ to $[v,X]$. In $\bar{G}_M$, the $i$th layer contains exactly the nodes of level
$i$. It is clear that $s$ is the only node in Layer 0. 
By the structure of~$G_M$, $\lev([v,B])$ is odd and $\lev([v,A])$ is even for all 
$v \in V$. Since breadth-first search (BFS) on $G_M$ with start node~$s$ finds shortest 
simple distances from~$s$ and not shortest strongly simple distances, BFS cannot be
used directly for the construction of $\bar{G}_M$. But we can modify BFS 
such that the modified breadth-first search (MBFS) finds shortest strongly 
simple distances.
Remember that for the construction of the $(l+1)$st layer, BFS needs only
to consider the nodes in Layer $l$, and to insert into the $(l+1)$st layer 
all nodes~$w$  which fulfill the following properties:
\begin{enumerate}
\item There is a node~$v$ in the $l$th layer with $(v,w)\in E$.
\item ${\rm Level}(w)$ has not been defined.
\end{enumerate}
With respect to finding strongly simple distances from~$s$, the construction of
the $(l+1)$st layer is a bit more difficult.
By the structure of $G_M$, the level of a free node $[w,B]$ is one and
the level of a non-free node~$[w,B]$ is
well-defined by the level of the unique node~$[v,A]$ with
$([v,A],[w,B])\in E_M$. Hence, the construction of odd layers is trivial.
For odd $l$, we shall describe the construction of the $(l+1)$st layer under 
the assumption that Layers~$0,1,2,\ldots,l$ are constructed.
It is clear that similar to BFS, MBFS has to insert into the $(l+1)$st layer 
all nodes $[w,A]\in V'$ which fulfill the following properties:
\begin{enumerate}
\item There is a node~$[v,B]$ in Layer $l$ with $([v,B],[w,A])
        \in E_M$, and there is a strongly simple path
        from~$s$ to~$[v,B]$ of length~$l$ which does not contain the node $[w,B]$.
\item ${\rm Level}([w,A])$ has not been defined.
\end{enumerate}
But these are not all nodes which MBFS has to insert into Layer $l+1$.
Consider the example described by Figure 2.
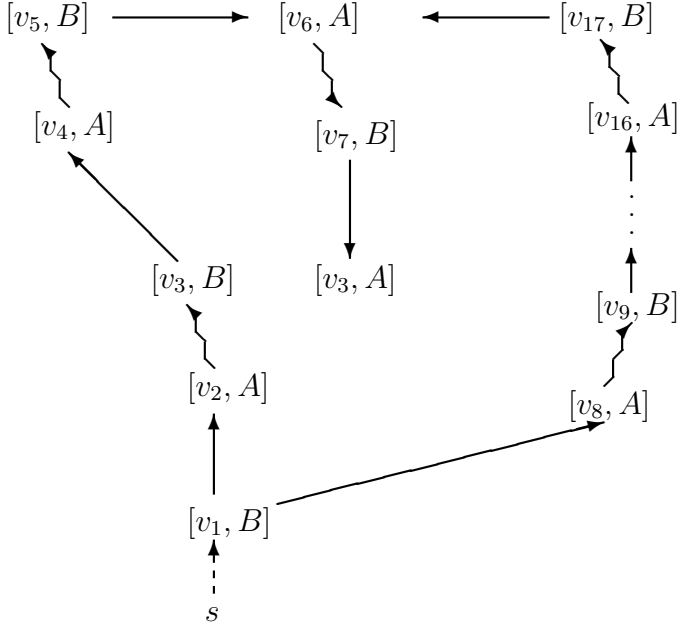
\begin{figure}[t]
\setlength{\unitlength}{0.0094in}%
\begin{picture}(345,346)(165,370)
\thicklines
\multiput(510,585)(0.00000,10.00000){3}{\makebox(0.5926,0.8889){\tenrm .}}
\put(260,570){\vector(-1, 1){ 60}}
\put(355,626){\vector( 0,-1){ 55}}
\multiput(335,690)(0.55556,-0.55556){10}{\makebox(0.5926,0.8889){\sevrm .}}
\put(340,685){\line( 0,-1){ 10}}
\multiput(340,675)(0.55556,-0.55556){10}{\makebox(0.5926,0.8889){\sevrm .}}
\put(345,670){\line( 0,-1){ 10}}
\multiput(345,660)(0.55556,-0.55556){10}{\makebox(0.5926,0.8889){\sevrm .}}
\put(350,655){\vector( 1,-1){0}}
\put(465,705){\vector(-1, 0){ 70}}
\put(224,705){\vector( 1, 0){ 75}}
\multiput(508,657)(-0.55556,0.55556){10}{\makebox(0.5926,0.8889){\sevrm .}}
\put(503,662){\line( 0, 1){ 10}}
\multiput(503,672)(-0.55556,0.55556){10}{\makebox(0.5926,0.8889){\sevrm .}}
\put(498,677){\line( 0, 1){ 10}}
\multiput(498,687)(-0.55556,0.55556){10}{\makebox(0.5926,0.8889){\sevrm .}}
\put(493,692){\vector(-1, 1){0}}
\put(510,614){\vector( 0, 1){ 25}}
\put(510,552){\vector( 0, 1){ 25}}
\multiput(495,500)(0.55556,0.55556){10}{\makebox(0.5926,0.8889){\sevrm .}}
\put(500,505){\line( 0, 1){ 10}}
\multiput(500,515)(0.55556,0.55556){10}{\makebox(0.5926,0.8889){\sevrm .}}
\put(505,520){\line( 0, 1){ 10}}
\multiput(505,530)(0.55556,0.55556){10}{\makebox(0.5926,0.8889){\sevrm .}}
\put(510,535){\vector( 1, 1){0}}
\put(315,435){\vector( 4, 1){180}}
\multiput(200,655)(-0.55556,0.55556){10}{\makebox(0.5926,0.8889){\sevrm .}}
\put(195,660){\line( 0, 1){ 10}}
\multiput(195,670)(-0.55556,0.55556){10}{\makebox(0.5926,0.8889){\sevrm .}}
\put(190,675){\line( 0, 1){ 10}}
\multiput(190,685)(-0.55556,0.55556){10}{\makebox(0.5926,0.8889){\sevrm .}}
\put(185,690){\vector(-1, 1){0}}
\multiput(280,510)(-0.55556,0.55556){10}{\makebox(0.5926,0.8889){\sevrm .}}
\put(275,515){\line( 0, 1){ 10}}
\multiput(275,525)(-0.55556,0.55556){10}{\makebox(0.5926,0.8889){\sevrm .}}
\put(270,530){\line( 0, 1){ 10}}
\multiput(270,540)(-0.55556,0.55556){10}{\makebox(0.5926,0.8889){\sevrm .}}
\put(265,545){\vector(-1, 1){0}}
\put(280,440){\vector( 0, 1){ 45}}
\multiput(280,385)(0.00000,8.57143){4}{\line( 0, 1){  4.286}}
\put(280,415){\vector( 0, 1){0}}
\put(275,370){\makebox(0,0)[lb]{\raisebox{0pt}[0pt][0pt]{\egtrm $s$}}}
\put(470,700){\makebox(0,0)[lb]{\raisebox{0pt}[0pt][0pt]{\egtrm $[v_{17},B]$}}}
\put(335,555){\makebox(0,0)[lb]{\raisebox{0pt}[0pt][0pt]{\egtrm $[v_3,A]$}}}
\put(335,635){\makebox(0,0)[lb]{\raisebox{0pt}[0pt][0pt]{\egtrm $[v_7,B]$}}}
\put(315,700){\makebox(0,0)[lb]{\raisebox{0pt}[0pt][0pt]{\egtrm $[v_6,A]$}}}
\put(485,645){\makebox(0,0)[lb]{\raisebox{0pt}[0pt][0pt]{\egtrm $[v_{16},A]$}}}
\put(490,540){\makebox(0,0)[lb]{\raisebox{0pt}[0pt][0pt]{\egtrm $[v_9,B]$}}}
\put(475,485){\makebox(0,0)[lb]{\raisebox{0pt}[0pt][0pt]{\egtrm $[v_8,A]$}}}
\put(165,700){\makebox(0,0)[lb]{\raisebox{0pt}[0pt][0pt]{\egtrm $[v_5,B]$}}}
\put(180,640){\makebox(0,0)[lb]{\raisebox{0pt}[0pt][0pt]{\egtrm $[v_4,A]$}}}
\put(245,555){\makebox(0,0)[lb]{\raisebox{0pt}[0pt][0pt]{\egtrm $[v_3,B]$}}}
\put(265,495){\makebox(0,0)[lb]{\raisebox{0pt}[0pt][0pt]{\egtrm $[v_2,A]$}}}
\put(265,420){\makebox(0,0)[lb]{\raisebox{0pt}[0pt][0pt]{\egtrm $[v_1,B]$}}}
\end{picture}
\caption{Further node which has to be inserted into Layer $l+1$.}
\end{figure}
Note that $\lev([v_7,B])=7$ but $\lev([v_3,A]) \not=8$, 
since the unique shortest strongly simple path from~$s$ to~$[v_7,B]$
contains $[v_3,B]$. The unique strongly simple path~$P$ from~$s$ to~$[v_3,A]$
has length~14. Hence, $\lev([v_3,A])=14$. Therefore, MBFS has to insert nodes 
$[w,A]\in V'$ into Layer $l+1$ for which there is a shortest 
strongly simple path~$P= s,[v_1,B], \ldots,[v_l,B],[w,A]$ with 
$\lev([v_l,B])<l$. For the treatment of these nodes and for the knowledge if
there is a strongly simple path from $s$ to $[v,B]$ of length $l$ which does
not contain the node $[w,B]$, the following notation is useful.

Let $T \subset V'$ such that $\lev([v,X])$ is defined for all $[v,X] \in  T$.
We denote by $\DOM(T)$ the set of those nodes 
$[u,B]\in V'\setminus T$ which satisfy:
\begin{itemize}
\item[a)] For all $[v,X] \in T$, all shortest strongly simple paths from~$s$ to
$[v,X]$ contain $[u,B]$,
\item[b)] $\lev([u,A])$ has not been defined, and
\item[c)] $\lev([w,B]) \leq \lev([u,B])$ for all $[w,B]\in V'$ satisfying a) and b).
\end{itemize}
If such a node $[u,B]$ does not exist then $\DOM(T)$ denotes the node $s$.
Furthermore, $\DOM(\{s\})$ denotes the node $s$.
We shall use $\DOM(T)$ only for subsets $T$ of $V'$ of size at most two.
Next we shall show that always $|\DOM(T)| = 1$. 
This will be a direct consequence of the following lemma.
\begin{lemma} \label{hk1}
Let $P = s,[v_1,B],[v_2,A], \ldots ,[v_l,Z]$ be any shortest strongly simple
path from $s$ to $[v_l,Z]$, i.e., $\mbox{level}([v_l,Z]) = l$. Let $[v_j,X]
\in P$ be a node with $\mbox{level}([v_j,\overline{X}]) \geq l$. Then $\lev([v_j,X]) = j$ 
and $\lev([v_i,Y]) < \lev([v_j,X])$ for all $i<j$.
\end{lemma}
{\bf Proof:} 
Since $[v_j,X]$ is the $j$th node on the strongly simple path $P$ it holds that
level$([v_j,X]) \leq j$.
Suppose $\lev([v_j,X]) < j$ and let $Q,[v_j,X]$ be any shortest strongly simple path
from $s$ to $[v_j,X]$. Let $P_2 := [v_{j+1},\overline{X}],[v_{j+2},X], \ldots,[v_l,Z]$. 
Note that all nodes on $P_2$ are not $M$-free. Let $R := Q,[v_j,X],P_2$.
By construction, $|R| < |P|$. Since $P$ is a shortest strongly simple path from $s$ 
to $[v_l,Z]$, the paths $Q$ and $P_2$ cannot be strongly disjoint. Let $[z,Y]$ be the
first node on $Q$ such that $[z,Y]$ or $[z,\overline{Y}]$ is on $P_2$. 
If $[z,Y]$ is on $P_2$ then $Y = A$. If $Y = B$ then the unique node $[x,A]$ with
$([x,A],[z,B]) \in E_M$ has to be the direct predecessor of the node $[z,B]$ on both 
paths $Q$ and $P_2$. This contradicts the choice of the node $[z,Y]$. 
If $[z,\overline{Y}]$ is on $P_2$ then also $Y = A$. If $Y = B$ then the node $[x,A]$ 
such that $([x,A],[z,B]) \in E_M$ has to be on $Q$ the direct predecessor of $[z,B]$ 
and $[x,B]$ has to be on $P_2$ the direct successor of $[z,A]$.
But this would also contradict the choice of the node $[z,Y]$. This shows
that in any case $Y = A$. Let
$$Q = Q_1,[z,A],Q_2 \quad \mbox{ and } \quad 
P_2 = \left\{ \begin{array}{ll}
               P_{21},[z,A],P_{22} & \mbox{if $[z,A] \in P_2$} \\
               P_{21},[z,B],P_{22} & \mbox{if $[z,B] \in P_2$.} 
              \end{array} \right.
$$
If $[z,A] \in P_2$, then $S := Q_1,[z,A],P_{22}$ would be a strongly 
simple path from $s$ to $[v_l,Z]$ shorter than $P$, a contradiction. 
Hence, $[z,B] \in P_2$. But then
$S := Q_1,[z,A],r(P_{21}),[v_j,\overline{X}]$
would be a strongly simple path from $s$ to $[v_j,\overline{X}]$ shorter than $l$. This
contradicts $\lev([v_j,\overline{X}]) \geq l$. Hence, level$([v_j,X]) = j$.

Since $[v_i,Y]$ is the $i$th node on $P$ it holds that $\lev([v_i,Y]) \leq i$.
Hence, $i < j$ implies $\lev([v_i,Y]) < \lev([v_j,X])$.
\QED

\medskip
The following lemma is a simple consequence of Lemma \ref{hk1}.
\begin{lemma} \label{hk1a}
Let $T \subset V'$, $T \not= \emptyset$ such that $\lev([v,X])$ is defined for all 
$[v,X] \in T$. Then the following statements hold true:
\begin{itemize}
\item[a)] 
$|\DOM(T)| = 1$.
\item[b)]
Let $\DOM(T) = [u,B]$. Then after the definition of $\lev([u,A])$, always
$\DOM(T) = \DOM([u,B])$.
\end{itemize}
\end{lemma}
{\bf Proof:} a) Assume that $|\DOM(T)| > 1$. Let $[u_1,B]$ and $[u_2,B]$ be two
distinct elements of $\DOM(T)$. By the consideration of any shortest strongly 
simple path from $s$ to $[v,X]$ for any $[v,X] \in T$, applying Lemma \ref{hk1}, we obtain 
$\lev([u_1,B]) < \lev([u_2,B])$ or $\lev([u_2,B]) < \lev([u_1,B])$. This
contradicts Part c) of the definition of $\DOM(T)$.

\medskip
\noindent
b)
After the definition of $\lev([u,A])$, $\lev(\DOM(T))$ decreases. Hence,
by Lemma \ref{hk1}, 
$\DOM(T)$ has to be below $[u,B]$ on all shortest strongly simple paths from
$s$ to $[v,X]$ for all $[v,X] \in T$. Hence, $\DOM([u,B])$ is on all shortest strongly 
simple paths from $s$ to $[v,X]$ for all $[v,X] \in T$. Assume that 
$\DOM(T) \not= \DOM([u,B])$. Then $\lev(\DOM(T)) > \lev(\DOM([u,B]))$. 
By the definition of $\DOM([u,B])$, there is a shortest strongly simple path $P_1,[u,B]$ 
from $s$ to $[u,B]$ 
which does not contain $\DOM(T)$. Hence, $P_1,[u,B]$ cannot be a subpath of a shortest 
strongly simple path from $s$ to $[v,X]$ for a node $[v,X] \in T$. Let $Q_1,[u,B],Q_2$ 
be any shortest strongly simple path from $s$ 
to $[v,X]$. Lemma \ref{hk1} implies that $|P_1,[u,B]| = |Q_1,[u,B]|$. Since 
$P_1,[u,B],Q_2$ is not a shortest strongly simple path, the paths $P_1$ and $Q_2$ are 
not strongly disjoint. Let $[z,A]$ be the first node on $P_1$ such that $[z,A]$ or $[z,B]$ 
is on $Q_2$. Exactly as in the proof of Lemma \ref{hk1}, we prove that such a node cannot 
exist. Hence, $\lev(\DOM(T)) \leq \lev(\DOM([u,B]))$. Therefore, $\DOM(T) = \DOM([u,B])$.
\QED

\medskip
Now we shall describe a phase in detail. Note that to each node $w \in V$ there correspond 
two different levels, namely $\lev([w,A])$ and $\lev([w,B])$. The {\em first level} of $w$ 
denoted by $\lev_1(w)$ is the smaller one of these two levels. The other level denoted by 
$\lev_2(w)$ is the {\em second level} of $w$.

In \cite{Bl1,Bl4}, we have constructed during the
$(l+1)$st phase exactly the shortest strongly simple paths of length $l+1$. Especially for
the treatment of the shortest strongly simple paths $P = s,[v_1,B], \ldots,[v_l,B],[w,A]$
with $\lev([v_l,B]) < l$, we have used some sophisticated data structures to achieve a 
certain time bound. In \cite{Bl4}, we have used the dynamic nearest common ancestor algorithm 
of \cite{Ga2} in combination with Fibonacci heaps \cite{FrTa} to get an
$O(m \log n)$ implementation of a phase. We have also described an alternative to the
use of Fibonacci heaps which uses two-dimensional arrays to get an $O(m + n^{3/2})$
implementation. A better way is to use an idea of Micali and Vazirani \cite{MV,Va1}. 
For all nodes $w$, they compute the first level during Phase $\lev_1(w)$ and the second level 
during Phase $\frac{1}{2}(\lev_1(w) + \lev_2(w) - 1) + 1$. We shall incorporate this idea into 
our framework.

Each phase separates into two parts. Both parts use breadth-first search.
The first level of each node $w$ is computed during Part 1 of Phase $\lev_1(w)$.
The second level of each node $w$ is computed during Part 2 of Phase
$\frac{1}{2}(\lev_1(w) + \lev_2(w) - 1)+1$. Before describing the two parts of a phase,
we shall investigate the structure of shortest strongly simple paths defining the first 
and the second level of a node.
First we characterize exactly those shortest strongly simple paths which define
the first level of a node to be $l+1$. 

\medskip
If $l$ is even then $\lev_1(w) = l+1$ for a node $w \in V$ iff for the unique node 
$v \in V$ such that $([v,A],[w,B]) \in E_M$ it holds that $\lev([v,A]) = \lev_1(v) = l$. 
Note that $\lev([v,B]) < l$ implies that $\lev([w,A]) < l$ such that $\lev([w,B])$ cannot 
be the first level of the node $w$. Each shortest strongly simple path from $s$ to the
node $[v,A]$ followed by the edge $([v,A],[w,B])$ is a shortest strongly simple path
from $s$ to $[w,B]$. Furthermore, there is no other shortest strongly simple path from
$s$ to $[w,B]$.

If $l$ is odd then $\lev_1(w) = l+1$ for a node $w \in V$ iff $\lev([w,A])$ and 
$\lev([w,B])$ are larger than $l$ and there is a node $[v,B]$ with $\lev([v,B]) = l$
and the edge $([v,B],[w,A])$ is in $E_M$. Exactly the shortest strongly simple paths
from $s$ to such a node $[v,B]$ followed by the edge $([v,B],[w,A])$ are
the shortest strongly simple paths from $s$ to $[w,A]$.

\medskip
Next, we shall characterize exactly those shortest strongly simple paths which define
the second level of a node to be $l+1$. 

\medskip
If $l$ is even then $\lev_2(w) = l+1$ for a node $w \in V$ iff $\lev([w,B]) > l$,
$\lev([w,A]) < l$, and for the unique node $v \in V$ such that $([v,A],[w,B]) \in E_M$ 
it holds that $\lev([v,A]) = l$. Note that the node $[w,A]$ cannot be on a strongly simple
path from $s$ to $[v,A]$ since the unique direct successor of $[w,A]$ has to be the node
$[v,B]$. With respect to the node $[v,A]$, two cases
are possible, $\lev([v,A]) = \lev_1(v)$ and $\lev([v,A]) = \lev_2(v)$. Exactly the shortest 
strongly simple paths from $s$ to the node $[v,A]$ followed by the edge $([v,A],[w,B])$
are the shortest strongly simple paths from $s$ to $[w,B]$.

If $l$ is odd and $\lev_2(w) = l+1$ for a node $w \in V$ then $\lev([w,A]) > l$ and 
$\lev([w,B]) < l$. With respect to a shortest strongly simple path $P$ from $s$ to $[w,A]$,
for the direct predecessor $[v,B]$ of $[w,A]$, exactly two situations are possible,
$\lev([v,B]) = l$ and $\lev([v,B]) < l$.

In the case that $\lev([v,B]) = l$ two subcases can happen, $\lev([v,B]) = \lev_1(v)$ and
$\lev([v,B]) = \lev_2(v)$. Exactly the shortest strongly simple paths $P$ from $s$ to such a 
node $[v,B]$ with the property that $[w,B]$ is not on $P$ followed by the edge 
$([v,B],[w,A])$ are the shortest strongly simple paths from $s$ to $[w,A]$.

If $\lev([v,B]) < l$ then consider any shortest strongly simple path\\ $P = P',[v,B],[w,A]$ from 
$s$ to $[w,A]$. Let $[x,B]$ be the last node on $P',[v,B]$ such that the length of the subpath
from $s$ to $[x,B]$ of $P$ is equal $\lev([x,B])$. Since $\lev([v,B]) < l = \lev([w,A]) - 1$,
the node $[x,B]$ has to be on $P'$. Since the level of the direct successor of $s$ on $P$ is
one, the node $[x,B]$ on $P'$ exists. 
Let $P' = P_1,[x,B],[y,A],P_2$. As we shall prove later,
$[y,A],P_2,[v,B],[w,A]$ has to be on the back-path of a shortest strongly simple path from $s$
to the node $[y,B]$.

\medskip
Next, we shall describe MBFS in detail.
At the beginning, $V' = \emptyset$ and $\bar{E}_M = \emptyset$.
In the first phase, the node $s$ is inserted into Layer 0. For each $M$-free node $w \in V$,
the node $[w,B]$ is inserted into Layer 1 and the edge $(s,[w,B])$ is inserted into
$\bar{E}_M$. Assume that $l > 0$ and Phase $l$ is finished. 
We shall give a detailed description of the two parts of Phase $l+1$.

\medskip
\noindent
{\em Part 1 of Phase $l+1$}

\medskip
If $l$ is even then MBFS adds for all nodes $[v,A]$ with $\lev([v,A]) = l = \lev_1(v)$ 
the unique edge $([v,A],[w,B]) \in E_M$ to $\bar{E}_M$ and inserts the node $[w,B]$ into 
the $(l+1)$st layer. Note that $\lev([v,A]) = \lev_2(v)$ implies that the first level of
node $w$ is already defined.

If $l$ is odd then MBFS considers all edges 
$([v,B],[w,A])$ with level$([v,B]) = l$ and $([v,B],[w,A])$ has not been considered 
during Part 2 of a previous phase. If the edge $([v,B],[w,A])$ has been considered during
Part 2 of a previous phase then $\lev_1(w)$ is already defined. Three cases are possible.

\medskip
\noindent
{\bf Case 1:} $\lev([w,A]) > l$ and $\lev([w,B]) > l$.

\medskip
MBFS inserts the node $[w,A]$ into the $(l+1)$st layer and adds the edge
$([v,B],[w,A])$ to $\bar{E}_M$.

\medskip
\noindent
{\bf Case 2:} $\lev([w,A]) > l$ and $\lev([w,B]) \leq l$.

\medskip
Then the first level of the node $w$ is already defined. But for the computation of the
second level of $w$ and some other nodes, the edge $([v,B],[w,A])$ is needed. Hence, MBFS 
inserts the pair
$([v,B],[w,B])$ into the set $E(k)$ where $k := \frac{1}{2}(\lev([v,B]) + \lev([w,B]))$.

\medskip
\noindent
{\bf Case 3:} $\lev([w,A]) \leq l$.

\medskip
Then the first level of the node $w$ is also defined. 
But the edge $([v,B],[w,A])$ might be necessary for the computation of the second level of 
some nodes. Hence, if $\lev([w,B])$ is already computed then the pair $([v,B],[w,B])$ is 
inserted into $E(k)$ where $k := \frac{1}{2}(\lev([v,B]) + \lev([w,B])$. Otherwise, this pair 
is inserted into $E(k)$ directly after the computation of $\lev([w,B])$.

\medskip
These are all cases. Next we shall describe the second part of Phase $l+1$.

\medskip
\noindent
{\em Part 2 of Phase $l+1$}

\medskip
During Part 2 of Phase $l+1$, $\lev_2(w)$ will be computed for all nodes $w$ with
$\frac{1}{2}(\lev_1(w) + \lev_2(w) - 1) = l$. 
MBFS considers all pairs of nodes $[x,Z],[y,Z]$ where $Z = A$ if $l$ is even and 
$Z = B$ if $l$ is odd such that
\begin{enumerate}
\item
$\lev([x,Z])$ and $\lev([y,Z])$ have been defined,
\item
$([x,Z],[y,\overline{Z}]) \in E_M$, and
\item
$\frac{1}{2}(\lev([x,Z]) + \lev([y,Z])) = l$.
\end{enumerate}
Note that these are exactly the pairs in $E(l)$.
Starting in $[x,Z]$ and $[y,Z]$, MBFS performs a breadth-first search on the back-paths 
of the current layered network until $\DOM(\{[x,Z],[y,Z]\})$ is reached.
All visited nodes $[u,X]$ such that $\lev([u,X])$ has not been defined
and $[u,X] \not= \DOM(\{[x,Z],[y,Z]\})$ are inserted into Layer $k$ where
$k = \lev([x,Z]) + \lev([y,Z]) + 1 - \lev([u,\overline{X}])$. 
The edges $([x,Z],[y,\overline{Z}])$ and $([y,Z],[x,\overline{Z}])$ are inserted into 
$\bar{E}_M$. Furthermore, MBFS adds the traversed edges which are not in $\bar{E}_M$ 
to $\bar{E}_M$.

\subsection{The correctness proof of MBFS}

We say that a path $P$ {\em is constructed} by MBFS if all edges on $P$ are inserted into
$\bar{E}_M$. We have to prove the following: 
\begin{enumerate}
\item
For all nodes $w \in V$, MBFS computes both levels $\lev_1(w)$ and $\lev_2(w)$ correctly.
\item
The layered network $\bar{G}_M = (V',\bar{E}_M)$ computed by MBFS contains all shortest
strongly simple paths from $s$ to $t$.
\end{enumerate}
The correctness of MBFS is a direct consequence of the following theorem.

\begin{theo} \label{hk4}
Let $w \in V$ and $\lev_1(w) = \lev([w,X])$. Then the following holds:
\vspace{-0.2cm}
\begin{itemize}
\item[a)]
MBFS defines $\lev([w,X])$ correctly during Part 1 of Phase $\lev([w,X])$. All 
shortest strongly simple paths from $s$ to $[w,X]$ have been constructed after the 
termination of Part 1 of Phase $\lev([w,X])$. 
\vspace{-0.2cm}
\item[b)]
Level$([w,\overline{X}])$ has been defined and all shortest strongly simple paths from $s$ to 
$[w,\overline{X}]$ have been constructed after the termination of Part 2 of Phase 
$\frac{1}{2}(\lev([w,A]) + \lev([w,B]) - 1) + 1$.
\end{itemize}
\end{theo}

\noindent
{\bf Proof:} 
We prove the theorem by induction on the number of performed phases. 
By construction, it is clear that the assertion of the theorem is fulfilled after the 
termination of Phase 1. Note that no second level of a node has to be computed during
Phase 1. Assume that $l > 0$ and that the assertion of the theorem is fulfilled after
the termination of Phase $l$. We shall prove that the assertion also holds
after the termination of Phase $l+1$.

First we shall prove that for all nodes with first level $l+1$, their first level is computed 
correctly during Part 1 of Phase $l+1$. Furthermore, the corresponding shortest strongly simple 
paths are constructed after termination of Part 1 of Phase $l+1$. No level of any other node
is determined during Part 1 of Phase $l+1$.

If the assertion of the theorem does not hold for a node $[u,B]$ then the assertion also
does not hold for the unique node $[v,A]$ with $([v,A],[u,B]) \in E_M$. Hence, if $l$ is
even, the induction hypothesis implies that the assertion is fulfilled after the termination
of Part 1 of Phase $l+1$. Therefore, we have only to consider the case that $l$ is odd.

Consider any node $u \in V$ with $\lev_1(u) = \lev([u,A]) = l+1$. Let $P$ be any shortest 
strongly simple path from $s$ to $[u,A]$. Let $[x,B]$ be the direct predecessor of $[u,A]$ 
on $P$. Since $\lev([u,B]) > \lev([u,A])$, the node $[u,B]$ cannot be a node on a shortest 
strongly simple path from $s$ to $[x,B]$. Hence, $\lev([x,B]) = \lev([u,A]) - 1$.
Since the assertion of the lemma has been maintained in previous phases, the node $[x,B]$
and the edge $([x,B],[u,A])$ have been considered during Part 1 of Phase $l+1$. Therefore,
the node $[u,A]$ is inserted into Layer $l+1$ and the edge $([x,B],[u,A])$ is inserted into 
$\bar{E}_M$. Since each shortest strongly simple path from $s$ to $[x,B]$ is contained in the 
current layered network, the path $P$ is constructed as well.
On the other hand, if we have a shortest strongly simple path $P$ from $s$ to a node $[x,B]$ 
of length $l$ and
an edge $([x,B],[u,A])$, where $\lev_1(u)$ is not defined, then the path $P,[u,A]$ is a shortest
strongly simple path from $s$ to $[u,A]$. Hence, $\lev_1(u) = l+1$. Therefore, only for nodes 
with first level $l+1$, the first level is computed during Part 1 of Phase $l+1$. 
This shows that the assertion is fulfilled after the termination of Part 1 of Phase $l+1$
such that Part a) of the theorem is proved.

\medskip
Next we shall show that the assertion of the theorem is fulfilled after the termination
of Part 2 of Phase $l+1$. It is clear by construction that during Part 2 of a phase only 
the second level of some nodes is computed. We have to prove that after the termination 
of Part 2 of Phase $l+1$, the following properties are fulfilled.
\begin{enumerate}
\item
For all nodes $[u,X]$ such that $\lev([u,X]) := k$ is computed during Part 2 of Phase $l+1$ 
it holds that $\frac{1}{2}(\lev([u,A]) + \lev([u,B]) - 1) = l$ and $\lev([u,X])$ is computed
correctly.
\item
For all nodes $u \in V$ with $\frac{1}{2}(\lev([u,A]) + \lev([u,B]) - 1) = l$, 
the second level of $u$ is computed correctly during Part 2 of Phase $l+1$ and
all shortest strongly simple paths from $s$ to $[u,X]$ with $\lev([u,X]) = \lev_2(u)$ have
been constructed after the termination of Part 2 of Phase $l+1$.
\end{enumerate}
First, we shall prove that the first property is fulfilled and then we shall consider the
second property. 

\medskip
Assume that the first property is not fulfilled. Consider the first situation such that
the first property is not fulfilled. Assume that this situation occurs during the backward 
search with respect to the pair $[x,Z][y,Z]$ of nodes because of the assignment 
$\lev([u,X]) := k$. Since a breadth-first search on the back-paths is performed and this is 
the first time that the first property is not fulfilled, $\lev([u,X])$ has to be larger than 
$k$. Furthermore, for each predecessor of $[u,X]$ on the back-path, its second level is computed
correctly. 
By the construction of the algorithm MBFS, there also holds $\lev([x,Z]) + \lev([y,Z]) = 2l$.
First we shall prove that $\frac{1}{2}(\lev([u,A]) + \lev([u,B]) - 1) = l$ and then that 
there is a shortest strongly simple path from $s$ to $[u,X]$ of length $k$.

By construction, $k = \lev([x,Z]) + \lev([y,Z]) + 1 - \lev([u,\overline{X}])$. 
Hence,
$$\lev([u,X]) + \lev([u,\overline{X}]) = \lev([x,Z] ) + \lev([y,Z]) + 1 = 2l + 1$$
and therefore,
$$\frac{1}{2}(\lev([u,A]) + \lev([u,B]) - 1) = l.$$
Assume that no shortest strongly simple path from $s$ to $[u,X]$ of length $k$ exists. 
By Lemma \ref{hk1}, on all shortest strongly simple paths which contain the node 
$[u,\overline{X}]$, the length of the subpath from $s$ to $[u,\overline{X}]$ is equal
$\lev([u,\overline{X}]$. Furthermore, since $\lev([u,X])$ is computed during the backward 
search, $[u,\overline{X}] \not= \DOM(\{[x,Z],[y,Z]\})$. Moreover, 
$\lev(\DOM(\{[x,Z],[y,Z]\})) < \lev([u,\overline{X}])$.

Since no shortest strongly simple path from $s$ to $[u,X]$ of length $k$ exists, the node
$[u,\overline{X}]$ has to be on the all shortest strongly simple paths from $s$ to the direct
predecessor of $[u,X]$ during the backward search. Each such strongly simple path consists of
a shortest strongly path from $s$ to $[x,Z]$ or $[y,Z]$ and the backpath of a shortest strongly 
simple path from $s$ to $[y,Z]$ and $[x,Z]$, respectively until the node $[u,X]$ is reached 
without the node $[u,X]$. Therefore, the subpath from $s$ to $[x,Z]$ and $[y,Z]$, 
respectively of these paths contains the node $[u,\overline{X}]$.
Since $[u,\overline{X}] \not= \DOM(\{[x,Z],[y,Z]\})$, there is a shortest strongly simple path 
from $s$ to at least one of the nodes in $\{[x,Z],[y,Z]\}$ which does not contain the node 
$[u,\overline{X}]$. W.l.o.g., let $R,[x,Z]$ be such a path.
Let $S = S_1,[u,\overline{X}],S_2,[y,Z]$ and $S' = S'_1,[u,\overline{X}],S'_2,[x,Z]$ be any 
shortest strongly simple paths from $s$ to $[y,Z]$ and to $[x,Z]$, respectively such that
the path $S_1,[u,\overline{X}],S_2,[y,Z],[x,\overline{Z}],r(S'_2)$ 
is strongly simple and constructed during the backward search. Consider the path
$$T := R,[x,Z],[y,\overline{Z}],r(S_2),[u,X].$$
By construction, $|T| = k$. Since $\lev([u,X]) > k$, the paths $R$ and $r(S_2)$ and hence, the
paths $R$ and $S_2$, are not strongly disjoint. Let $[c,A]$ be the first node on $R$ 
such that one of the nodes $[c,A]$ or $[c,B]$ is on $S_2$. Let
$$R = R_1,[c,A],R_2 \quad \mbox{ and } \quad 
S_2 = \left\{ \begin{array}{ll}
               S_{21},[c,A],S_{22} & \mbox{if $[c,A] \in S_2$} \\
               S_{21},[c,B],S_{22} & \mbox{if $[c,B] \in S_2$.} 
              \end{array} \right.
$$
If $[c,B]$ is on $S_2$ then $R_1,[c,A],r(S_{21}),[u,X]$ would be a strongly simple path from
$s$ to $[u,X]$ shorter than $\lev([u,X])$, a contradiction. Hence, $[c,A]$ is a node on $S_2$.
By construction, $R_1,[c,A],S_{22},[y,Z]$ is strongly simple. Hence, 
$$|R_1,[c,A]| \geq |S_1,[u,\overline{X}],S_{21},[c,A]|.$$
If this inequality is strict then $|S_1,[u,\overline{X}],S_{21},[c,A],R_2,[x,Z]| < \lev([x,Z])$.
Hence, the paths $S_1,[u,\overline{X}],S_{21}$ and $R_2$ are not strongly disjoint. 
Let $[d,A]$ be the first node on $S_1,[u,\overline{X}],S_{21}$ such that one of the nodes
$[d,A]$ or $[d,B]$ is on $R_2$. If $[d,A]$ is on $R_2$ then a strongly simple path from $s$
to $[x,Z]$ shorter than $\lev([x,Z])$ can easily be constructed, a contradiction. Hence, $[d,B]$ is 
a node on $R_2$. Let $R_2 = R_{21},[d,B],R_{22}$.  We distinguish two cases.

\medskip
\noindent
{\em Case 1:} $[d,A]$ is on $S_{21}$.

\medskip
Let $S_{21} = S_{211},[d,A],S_{212}$. Consider the path 
$$T := R_1,[c,A],R_{21},[d,B],r(S_{211}),[u,X].$$
By construction, $T$ is a path from $s$ to $[u,X]$ shorter than $\lev([u,X]$.
By the choice of the nodes $[c,A]$ and $[d,A]$, the path $T$ is also strongly simple, a 
contradiction. Hence, Case 1 cannot happen.

\medskip
\noindent
{\em Case 2:} $[d,A]$ is on $S_1$.

\medskip
Let $S_1 = S_{11},[d,A],S_{12}$. Consider the path
$$T := S_{11},[d,A],r(R_{21}),[c,B],r(S_{21}),[u,X].$$
Note that $|T| < \lev([u,X])$. Hence, $T$ cannot be strongly simple. By the choice of the nodes
$[c,A]$ and $[d,A]$, the paths $R_{21}$ and $S_{21}$ are not strongly disjoint. Let
$[h,A]$ be the first node on $R_{21}$ such that one of the nodes $[h,A]$ or $[h,B]$ is a node on 
$S_{21}$. 
If $[h,A]$ is on $S_{21}$ then a strongly simple path from $s$ to $[u,X]$ shorter than $\lev([u,X])$ 
can easily be constructed, a contradiction. Hence, $[h,B]$ is a node on $S_{21}$. 
Let $R_{21} = R_{211},[h,A],R_{212}$ and $S_{21} = S_{211},[h,B],S_{212}$. Consider the path
$$T := R_1,[c,A],R_{211},[h,A],r(S_{211}),[u,X].$$
By construction, $T$ is a path from $s$ to $[u,X]$ shorter than $\lev([u,X])$.
By the choice of the nodes $[c,A]$ and $[h,A]$, the path $T$ is also strongly simple, a contradiction. 
Hence, also Case 2 cannot happen.

\medskip
Altogether, we have proved $|R_1,[c,A]| = |S_1,[u,\overline{X}],S_{21},[c,A]|$. Consider the path
$$T := R_1,[c,A],S_{22},[y,Z],[x,\overline{Z}],r(S'_2),[u,X].$$
By construction, $|T| = k$. Since $\lev([u,X]) > k$, the path $T$ cannot be strongly simple. By
construction, the paths $R_1$ and $r(S'_2)$ are not strongly disjoint. Let $[d,A]$ be the first
node on $R_1$ such that one of the nodes $[d,A]$ or $[d,B]$ is on $r(S'_2)$. 
If $[d,A]$ is on $r(S'_2)$ then a strongly simple path from $s$ to $[u,X]$ shorter than $\lev([u,X])$ 
can easily be constructed, a contradiction. Hence, $[d,B]$ is a node on $r(S'_2)$ and hence, $[d,A]$
is on $S'_2$.  Let
$R_1 = R_{11},[d,A],R_{12}$ and $S'_2 = S'_{21},[d,A],S'_{22}$. Consider the path
$$T := R_{11},[d,A],S'_{22},[x,Z],[y,\overline{Z}],r(S_2),[u,X].$$ 
By construction, the path $T$ is strongly simple. Since $S'$ is a shortest strongly simple
path from $s$ to $[x,Z]$ it holds that $|R_{11},[d,A]| \geq |S'_1,[u,X],S'_{21},[d,A]|$.
In the same way as we have proved $|R_1,[c,A]| = |S_1,[u,\overline{X}],S_{21},[c,A]|$, we prove 
$|R_{11},[d,A]| = |S'_1,[u,X],S'_{21},[d,A]|$ . Hence, $|T| = k$. This contradicts our assumption 
that $\lev([u,X]) > k$. 
Altogether, the first property is proved.

\medskip
Assume that the second property is not fulfilled. Let $u \in V$ be a node with smallest second
level such that the second property is not fulfilled with respect to the node $u$. This means 
that
\begin{enumerate}
\item
$\frac{1}{2}(\lev([u,A]) + \lev([u,B]) - 1) = l$ and there is a shortest strongly simple path 
$P = P',[v,\overline{X}],[u,X]$ with $\lev_2(u) = \lev([u,X])$ but 
\item
$P$ has not been constructed after the termination of Part 2 of Phase $l+1$ or 
$\lev_2(u)$ has not been computed correctly. 
\end{enumerate}
With respect to $\lev([v,\overline{X}])$, the situations $\lev([v,\overline{X}]) = 
\lev([u,X]) - 1$ and $\lev([v,\overline{X}]) < \lev([u,X]) - 1$ are possible. We shall investigate 
both cases one after the other.

\medskip
\noindent
{\em Case 1:} $\lev([v,\overline{X}]) = \lev([u,X]) - 1$.

\medskip
Two subcases are to consider, $\lev([v,\overline{X}]) = \lev_1(v)$ and
$\lev([v,\overline{X}]) = \lev_2(v)$.

\medskip
\noindent
{\em Subcase 1.1:} $\lev([v,\overline{X}]) = \lev_1(v)$.

\medskip
First we shall prove that $\lev([v,\overline{X}]) = \lev([u,\overline{X}]) = l$. 
If $\lev([v,\overline{X}]) > l$ then $\lev([u,X]) = \lev([v,\overline{X}]) + 1$ and
$\lev([u,A]) + \lev([u,B]) = 2l+1$ imply $\lev([u,\overline{X}]) < l$.
Let $R$ be any shortest strongly simple path from $s$ to
$[u,\overline{X}]$. Since $\lev_1(v) > l$, the nodes $[v,A]$ and $[v,B]$ cannot be on $R$. 
Hence, $R,[v,X]$ would be a strongly simple path from $s$ to $[v,X]$ shorter than 
$\lev_1(v)$, a contradiction. Furthermore, $\lev_2(u) = \lev([v,\overline{X}]) + 1$ and
$\lev([u,A]) + \lev([u,B]) = 2l+1$ imply $\lev([v,\overline{X}]) = l$ and also
$\lev([u,\overline{X}]) = l$. Hence, during Part 1 of Phase $l+1$, Case 2 applies such
that the pair $[u,\overline{X}],[v,\overline{X}]$ is inserted into $E(l)$. Therefore,
during Part 2 of Phase $l+1$, the backward search with respect to the pair 
$[u,\overline{X}],[v,\overline{X}]$ has been performed.

By the induction hypothesis, at the beginning of Part 2 of Phase $l+1$, all shortest strongly
simple paths from $s$ to $[v,\overline{X}]$ and all shortest strongly simple paths from $s$
to $[u,\overline{X}]$ are constructed. Since during Part 2 of Phase $l+1$, the backward search with 
respect to the pair $[u,\overline{X}],[v,\overline{X}]$ is performed, the edges
$([v,\overline{X}],[u,X])$ and $([u,\overline{X}],[v,X])$ are inserted into the layered graph. 
Hence, after the termination of Part 2 of Phase $l+1$, the path 
$P = s, \ldots,[v,\overline{X}],[u,X]$ is constructed. By construction of the algorithm MBFS,
$\lev([u,X])$ has been computed correctly as well. This contradicts our assumption such that
Subcase 1.1 cannot happen.

\medskip
\noindent
{\em Subcase 1.2:} $\lev([v,\overline{X}]) = \lev_2(v)$.

\medskip
Let $R = R',[u,\overline{X}]$ be any shortest strongly simple path from $s$ to 
$[u,\overline{X}]$. Consider the path $Q := R',[u,\overline{X}],[v,X]$. 

If $Q$ is a shortest strongly simple path from $s$ to $[v,X]$ then 
$\lev([v,A]) + \lev([v,B]) = 2l+1$. 
By the choice of the node $u$, all shortest strongly simple paths from $s$ to 
$[v,\overline{X}]$ has been constructed. By the induction hypothesis, $R$ is also constructed. 
Hence, during the backward search which constructs the path $s,P',[v,\overline{X}]$,
the edge $([v,\overline{X}],[u,X])$ is added to the layered network. Hence, the path
$P$ is also constructed, a contradiction.
Therefore, $Q$ is not a shortest strongly simple path from $s$ to $[v,X]$. 

If $Q$ is strongly simple
then $\lev([v,X]) < |Q|$. Hence, $\lev([v,A]) + \lev([v,B]) < \lev([u,A]) + \lev([u,B]) = 2l+1$. 
By the induction hypothesis, $\lev([v,\overline{X}])$ has been computed and all shortest strongly 
simple paths from $s$ to $[v,\overline{X}]$ have been constructed. Furthermore, 
$\lev_1(u) = \lev([u,\overline{X}]) \leq l$. Hence, by the induction hypothesis, all shortest 
strongly simple paths from $s$ to $[u,\overline{X}]$ have been constructed and the edge 
$([u,\overline{X}],[v,X])$ has been considered during Part 1 of Phase
$\lev([u,\overline{X}]) + 1$. 
Therefore, during Part 2 of a Phase $k$ where $k \leq l+1$, the backward search has been performed
with respect to the pair $[v,\overline{X}],[u,\overline{X}]$ and the edge 
$([v,\overline{X}],[u,X])$ has been inserted into the layered network such that the path $P$
has been constructed after the termination of Part 2 of Phase $k$, a contradiction.

Hence, the path $Q = R',[u,\overline{X}],[v,X]$ is not strongly simple. Since $R',[u,\overline{X}]$
is strongly simple, the node $[v,X]$ or the node $[v,\overline{X}]$ has to be on the path $R'$. Let
$$
R' = \left\{ \begin{array}{ll}
               R'_1,[v,X],R'_2 & \mbox{if $[v,X] \in R'$} \\
               R'_1,[v,\overline{X}],R'_2 & \mbox{if $[v,\overline{X}] \in R'$.} 
              \end{array} \right.
$$
If $[v,X]$ is on $R'$ then $\lev([v,X]) < |Q|$. Exactly as in the case ``$Q$ is strongly simple but
$|Q| > \lev([v,X])$'', we prove that the path $P$ has been constructed after the termination of 
Part 2 of Phase $l+1$ getting a contradiction. Hence, $[v,\overline{X}]$ is on $R'$. But then, 
$R'_1,[v,\overline{X}],[u,X]$ would be a strongly simple path from $s$ to $[u,X]$ shorter than 
$\lev([u,X])$, a contradiction. Therefore, Subcase 1.2 and hence, Case 1 cannot happen.

\medskip
\noindent
{\em Case 2:} $\lev([v,\overline{X}]) < \lev([u,X]) - 1$.

\medskip
Let $Q,[u,\overline{X}]$ be any shortest strongly simple path from $s$ to $[u,\overline{X}]$. 
Let $[x,B]$ be the last node on $P',[v,\overline{X}]$ such that the length of the subpath from $s$ 
to $[x,B]$ of $P$ is equal to $\lev([x,B])$. Let $P = P_1,[x,B],[y,A],P_2,[v,\overline{X}],[u,X]$,
let $R := [u,\overline{X}],[v,X],r(P_2),[y,B]$ and let 
$S := Q,[u,\overline{X}],[v,X],r(P_2),[y,B]$.

First we shall show that for all nodes $[z,Y]$ on $R$, the level of $[z,Y]$ is not larger than
the length of the subpath $S([z,Y]) := Q,[u,\overline{X}],r(P_{21}),[z,Y]$ of $S$. Then we shall use 
this fact to prove that all edges on the path $R$ are contained in the current layered graph. 
Since $\lev([y,B]) \leq |S([y,B])|$, the pair $[x,B],[y,B]$ has been considered during Part 2
of a Phase $k$ where $k \leq l+1$. Hence, the path $P$ is constructed after the execution of 
Part 2 of Phase $k$, a contradiction.

To prove $\lev([z,Y]) \leq |S([z,Y])|$ assume that $\lev([z,Y]) > |S([z,Y])|$ for a node $[z,Y]$
on $R$. Then $S([z,Y])$ cannot be strongly simple such that $Q$ and $P' := r(P_{21}),[z,Y]$ are
not strongly disjoint. Let $[p,A]$ be the first node on $Q$ such that one of the nodes
$[p,A]$ or $[p,B]$ is a node on $P'$. Let 
$$Q = Q_1,[p,A],Q_2 \quad \mbox{ and } \quad 
P' = \left\{ \begin{array}{ll}
               P'_1,[p,A],P'_2 & \mbox{if $[p,A] \in P'$} \\
               P'_1,[p,B],P'_2 & \mbox{if $[p,B] \in P'$.} 
              \end{array} \right.
$$
If $[p,A] \in P'$ then $T := Q_1,[p,A],P'_2$ would be a strongly simple path from $s$ to $[z,Y]$ 
shorter than $|S([z,Y])|$, a contradiction. Hence, $[p,B] \in P'$. But then,
$T := Q_1,[p,A],r(P'_1),[u,X]$ would be a strongly simple path from $s$ to $[u,X]$ shorter than
$\lev([u,X])$, a contradiction. Altogether, we have proved that $\lev([z,Y]) \leq |S([z,Y])|$.

Assume that an edge $([z,Y],[z',\overline{Y}])$ is not contained in the current layered network
after the execution of Part 2 of Phase $l+1$. Since $\lev([z,Y]) \leq |S([z,Y])|$, we obtain
$\lev([z',Y]) + \lev([z,Y]) \leq \lev([u,X]) + \lev([u,\overline{X}]) - 1 = 2l$. Hence,
$\frac{1}{2}(\lev([z',Y]) + \lev([z,Y])) \leq l$. Therefore, the pair $[z',Y][z,Y]$ has been
considered and the edge $([z,Y],[z',\overline{Y}])$ has been inserted into the current layered 
graph during Part 2 of a Phase $k$ where $k \leq l+1$, a contradiction. Altogether we have shown 
that the path $P$ is constructed after the execution of Part 2 of Phase $l+1$.
Moreover, since the pair $[x,B],[y,B]$ has been considered during Part 2 of Phase $l+1$,
$\lev([u,X])$ has been computed correctly because of the consideration of this pair of nodes.
This shows that Case 2 cannot happen as well. This proves the second property.

\medskip
Altogether, the theorem is proved.
\QED

\subsection{An implementation of an entire phase}

First we shall describe the implementation of MBFS, and then we shall show how to 
combine MBFS and MDFS to get an implementation of an entire phase.

Obviously, Part 1 of all phases can be implemented in such a way that the 
total time is bounded by $O(m + n)$. For the implementation of Part 2 of all 
phases, we have to describe how to implement the search on the back-paths,
starting in $[v,Z]$ and $[w,Z]$, until $\DOM(\{[v,Z],[w,Z]\})$ is reached.
Most importantly, since we do not know $\DOM(\{[v,Z],[w,Z]\})$ in advance, 
meaning that $\DOM(\{[v,Z],[w,Z]\})$ has to be computed simultaneously, we have to 
take care that the search does not continue beyond $\DOM(\{[v,Z],[w,Z]\})$.
Note that by Lemma \ref{hk1}, the subpaths from $s$ to $\DOM(\{[v,Z],[w,Z]\})$
are always shortest strongly simple paths from $s$ to $\DOM\{[v,Z],[w,Z]\}$.
Hence, we can perform a breadth-first search on the back-paths until the current 
level contains exactly one node. By Lemma \ref{hk1}, this node has to be 
$\DOM(\{[v,Z],[w,Z]\})$. 

With respect to the efficiency, at the moment when the search meets a node
$[u,X]$ for which $\lev([u,X]) = \lev_2(u)$ has been defined, we have to compute efficiently
the next node of the search having the property that its level is not defined. 
By definition, this node is $\DOM([u,X])$. As a consequence of Lemma \ref{hk1a}, we can 
maintain these nodes by a data structure for disjoint set union such that for the computation 
of $\DOM([u,X])$
one {\em find\/} operation would suffice. In that case, an extensible edge would be stored.
Using disjoint set union \cite{Ta}, the total time can be bounded to be $O((m+n)\alpha(m,n))$ 
where $\alpha$ is the inverse Ackermann function. Using incremental tree set union 
\cite{GaTa}, we obtain a total time bound of $O(m+n)$ for the computation of 
the next node such that its level is not defined.

The levels of the nodes $[v,Z]$ and $[w,Z]$ have not to be equal. If the level of the 
two nodes are different then we start the breadth-first search at the node with 
larger level. We always continue the search at the nodes with largest level
until all nodes in the front of the search are on the same level. But we do not need
to perform a precise breadth-first search. Hence, we can organize the search in the following way:
\begin{enumerate}
\item
In the front of the search, continue the search always in a node $[u,X]$ such that
there is another node $[p,Y]$ in the front of the search with $\lev([p,Y]) \leq \lev([u,X])$.
\item
If the front of the search contains exactly one node then stop the search.
\end{enumerate}
Next, we shall combine MBFS and MDFS for the implementation of an entire phase.
Knowing $\bar{G}_M$, a maximal set of up to $s$ and $t$ pairwise disjoint shortest
strongly simple paths from $s$ to $t$ in $\bar{G}_M$ can be computed using MDFS in $O(m + n)$ 
time. Every time, a strongly simple path~$P$ from~$s$ to~$t$ is found, all
nodes $[v,A]$, $[v,B]$ with $[v,A]\in P$ or $[v,B]\in P$ and all incident edges
are deleted from~$\bar{G}_M$. If a node gets zero indegree or zero
outdegree, then also this node, and all incident edges, are deleted.
Altogether, we have obtained the following theorem.

\begin{theo}
A maximum matching in a general graph $G=(V,E)$ can be computed in
$O(\sqrt{n} (m + n))$ time and $O(m+n)$ space, where 
$\left|V\right| = n$ and $\left| E\right| = m$.
\end{theo}

\section{The primal-dual method for the \\maximum weighted matching problem}

Let $G =(V,E)$ be an undirected graph. 
If we associate with each edge $(i,j) \in E$ a weight $w(i,j) > 0$ then we
obtain a {\em weighted undirected graph\/} $G = (V,E,w)$.
The weight $w(M)$ of a matching $M$ is the sum of the weights of the edges in
$M$. A matching $M \subseteq E$ has {\em maximum weight\/} iff  $\sum_{(i,j) 
\in M'}w(i,j) \leq \sum_{(i,j) \in M}w(i,j)$ for all matchings
$M' \subseteq E$.
Given a weighted undirected graph $G = (V,E,w)$, the {\em maximum weighted
matching problem\/} is finding a matching $M \subseteq E$ of maximum weight.
Our goal is to develop a method for the computation of a maximum weighted 
matching in a given weighted undirected graph.

\subsection{The description of the method}

Let $G = (V,E,w)$ be a weighted undirected graph. Let ${\cal F} = \{E_1,E_2, 
\ldots , E_r\}$, $E_i \subseteq E$ be a family of pairwise distinct subsets of 
$E$. With each node $i \in V$ we associate a {\em node weight\/} $\pi(i) 
\geq 0$. Furthermore, with each edge set $E_l \in \cal F$, we associate a 
{\em set weight\/} $\mu(E_l) \geq 0$. These new variables are called {\em dual 
variables}.
The values of the dual variables are treated in such a way that the following 
invariant is always maintained.
\begin{itemize}
\item
$w(i,j) \leq \pi(i) + \pi(j) + \sum_{(i,j) \in E_l} \mu(E_l)$ for all $(i,j) \in E$.
\end{itemize}
The right side of this inequality is the {\em dual weight $d(i,j)$} of
the edge $(i,j) \in E$.
We define the {\em dual weight $d(M)$} of a matching $M$ by
$$d(M) := \sum_{(i,j) \in M} d(i,j).$$
Note that because of the invariant, always $w(M) \leq d(M)$ for all matchings 
$M \subseteq E$.  
With respect to an arbitrary matching $M \subseteq E$, the maximal possible 
contribution of the node weight $\pi(i)$ to $d(M)$ is $\pi(i)$ since $i$ is 
adjacent to at most one edge in $M$. Note that $|E_l \cap M| \leq c(E_l)$ where
$c(E_l)$ is the size of a maximum cardinality matching with respect to $E_l$. 
Hence, the maximal possible contribution of the set weight $\mu(E_l)$ to $d(M)$ is
$c(E_l)\mu(E_l)$. Hence,
$$\sum_{i \in V}\pi(i) + \sum_{E_l \in {\cal F}}c(E_l)\mu(E_l)$$
is always an upper bound for the dual weight of any matching of $G$.
Therefore, with respect to a matching $M$, 
$$w(M) = \sum_{i \in V}\pi(i) + \sum_{E_l \in {\cal F}}c(E_l)\mu(E_l)$$ 
implies that the matching $M$ has maximum weight.
The question is now: When with respect to a matching $M$, this equation holds?

The right side of the equation contains for each edge in $M$ its dual weight.
Since the dual weight of an edge is at least as large as its weight, we obtain 
the necessary condition $d(i,j) = w(i,j)$ for all edges $(i,j) \in M$.
Since all summands in both sums of the right side of the equation are non-negative, 
the node weight $\pi(i)$ has to be zero for all $M$-free nodes $i \in V$.
Furthermore, for all $E_l$ such that $|M \cap E_l| < c(E_l)$, the set weight
$\mu(E_l)$ has also to be zero. On the other hand, these conditions are fulfilled with
respect to a matching $M$ for which $w(M) = d(M)$. Altogether, we have obtained the 
following necessary and sufficient {\em optimality conditions\/}:
\begin{enumerate}
\item
$r(i,j) := d(i,j) - w(i,j) = 0$ for all $(i,j) \in M$,
\item
$\pi(i) = 0$ for all $M$-free nodes $i \in V$, and
\item 
$\mu(E_l) > 0$ for $E_l \in \cal F$ implies $|E_l \cap M| = c(E_l)$.
\end{enumerate}
The value $r(i,j)$ is called the {\em reduced cost\/} of the edge $(i,j)$. Because of
the invariant which is maintained, the reduced cost of an edge is always non-negative.

\medskip
The primal-dual method for the weighted matching problem can be separated into 
rounds. The input of every round will be a matching $M$, a set $\cal F$ of pairwise 
distinct subsets of $E$ such that $\mu(E_l) > 0$ for all $E_l \in \cal F$, and values 
for the dual variables which fulfill the first and third optimality conditions with 
respect to the matching $M$. The second optimality condition can be unsatisfied with 
respect to some free nodes. Our goal within a round is to modify $M$ and the 
values of the dual variables such that Conditions 1 and 3 remain valid
and the number of nodes violating Condition 2 is strictly decreased.

A round divides into two steps, the {\em search step\/} and the
{\em extension step}. The search step tries to improve
the current matching by finding an augmenting path $P$ such that the number
of free nodes with node weight larger than zero can be decreased by the 
augmentation of $P$. Since Condition 1 has to be maintained, the search step
can only be performed on edges with reduced cost zero. If the search step cannot
improve the current matching then the extension step
changes the values of some dual variables using an appropriate value $\delta$. 
The extension step can decrease the reduced cost of some edges to zero. Hence, 
it is possible that the next search step will find an augmenting path.

During the search step, we use MDFS. Hence, we define with respect to the current 
matching $M$ the weighted directed bipartite graph $G_M := (V',E_M,w)$ where
\begin{eqnarray*}
V' &:=& \left\{[v,A],[v,B] \mid v\in V\right\}
       \cup \{s,t\} \qquad s,t\not\in V,\ s\not=t \\
E_M&:=& \left\{([v,A],[w,B]),([w,A],[v,B]) \mid (v,w)\in M\right\} \\
  & & \cup \left\{([x,B],[y,A]),([y,B],[x,A]) \mid (x,y)\in E\setminus M\right\} \\
  & & \cup \left\{(s,[v,B]), ([v,A],t) \mid v\in V \mbox{ is $M$-free}\right\}.
\end{eqnarray*}
This means that $E_M$ contains for each edge in $E$ two copies.
Both copies of the edge $(i,j)$ obtain weight $w(i,j)$, dual weight $d(i,j)$, and hence, 
reduced cost $r(i,j)$.
We arrange that edges with tail $s$ or head $t$ have always reduced cost zero. 
According to Condition 1,
it is only allowed to consider augmenting paths where all edges on these
paths have reduced cost zero. Hence, the input graph $G^*_M = (V',E^*_M,w)$ of the
search step will be the subgraph of $G_M$ containing exactly those edges in $E_M$ 
having reduced cost zero; i.e., 
$$E^*_M := \{([i,X],[j,\overline{X}]) \in E_M \mid r(i,j) = 0\} \cup 
\{(s,[i,B]),([i,A],t) \mid i \mbox{ $M$-free}\}.$$
By definition, for each edge $(i,j) \in M$ it holds that 
$([i,A],[j,B]), ([j,A],[i,B]) \in E^*_M$. We start with the empty matching $\emptyset$ 
and define the graph $G_{\emptyset} = (V',E_{\emptyset},w)$ as described above.
Let $W := \max_{(i,j) \in E} w(i,j)$. We initialize all node weights $\pi(i)$
by $\frac{W}{2}$. Altogether, we obtain the input graph 
$G^*_{\emptyset} = (V', E^*_{\emptyset},w)$ for the first search step where
$$\begin{array}{ll}
E^*_{\emptyset} = & \{([i,B],[j,A]), ([j,B],[i,A]) \mid ([i,B],[j,A]) \in 
E_{\emptyset}, \mbox{ } w(i,j) = W\}\\ 
     & \cup \{(s,[i,B]),([i,A],t)  \mid i \in V\}.
\end{array}$$
At the beginning, the family ${\cal F}$ of subsets of $E$ will be empty such 
that no set weight has to be defined. 
During the execution of the algorithm, the needed elements of ${\cal F}$ and the corresponding 
set weights will be defined. 
Whenever to some edge set $E_l \not\in \cal F$ a value
$\mu(E_l) > 0$ is associated, $E_l$ enters $\cal F$. 
As soon as $\mu(E_l)$ gets to be zero, the set $E_l$ is removed from $\cal F$.

A search step terminates with a matching $M$, a weighted directed graph
$G_M = (V',E_M,w)$, a current subgraph $G^*_M = (V',E^*_M,w)$ of $G_M$ such that
no $M$-augmenting path $P$ is contained in $G^*_M$, a current set $\cal F$ where
the set weight of each set in $\cal F$ is positive, and values 
for the dual variables. This is the input of the next extension step.

For the treatment of the extension step consider the expanded MDFS-tree 
$T_{exp}$, computed by the last MDFS on $G^*_M$. Note that this MDFS was 
unsuccessful; i.e., no path from the start node $s$ to the target node $t$ has been 
found. The goal is to add edges to $T_{exp}$ such that 
possibly an augmenting path can be found. Therefore, we have to decrease the 
reduced cost of edges with positive reduced cost. Such an edge $(i,j)$ has to 
be in $E \setminus M$ and $[i,B]$ has to be in $T_{exp}$. Let
$$V_A := \{[i,A] \mid i \in V\} \quad \mbox{ and } \quad 
V_B := \{[i,B] \mid i \in V\}.$$
Furthermore, let
$$A_T := V_A \cap T_{exp} \quad \mbox{ and for a moment } \quad 
B_T := V_B \cap T_{exp}.$$
Later, we shall see that according to the optimality
conditions which we have to maintain, some nodes in $V_B \cap T_{exp}$ are
not allowed to be a node in $B_T$. Hence, we shall modify the definition of
$B_T$ by removing exactly these nodes.

The idea is to decrease the reduced cost $r(i,j)$ of all edges $(i,j)$ with 
positive reduced cost and $[i,B] \in B_T$ by an appropriate value $\delta$.
This is done by decreasing the node weight $\pi(i)$ by $\delta$ for all nodes 
$i$ with $[i,B] \in B_T$.
As a consequence of the decrease of the node weight $\pi(i)$, the reduced cost
of each edge $e$ in $G_M^*$ with end node $[i,A]$ or $[i,B]$ becomes negative. 
Because of the invariant maintained by the method, such a reduced cost has to be increased 
until its value is zero again. We distinguish two cases:
\begin{enumerate}
\item
The other end node of $e$ is in $A_T$ but not in $B_T$.
\item
The other end node of $e$ is in $B_T$.
\end{enumerate}
If Case 1 is fulfilled then we can increase the reduced cost of the edge $e$ by 
increasing the node weight of the other end node of $e$ by $\delta$; i.e.,
we increase $\pi(j)$ by $\delta$ for all nodes $j$ such that $[j,A] \in A_T$
and $[j,B] \not\in B_T$.
Note that increasing the node weight $\pi(j)$ of a node $j$ increases the reduced 
cost of each edge $(i,j)$.
If $(i,j) \in E \setminus M$ then $r(i,j)$ has only to be non-negative. Since
$r(i,j)$ was non-negative before the increase of $\pi(j)$, it is also non-negative after 
the increase. If $(i,j) \in M$ then $r(i,j)$ has to be zero. Note that $[j,A] \in A_T$
implies for the unique node $[i,B]$ with $(i,j) \in M$ that $[i,B] \in B_T$.
Hence, $\pi(i)$ has been decreased by the same value such that $r(i,j) = 0$ 
before the change of the dual variables implies that $r(i,j) = 0$ after the
change.

If Case 2 is fulfilled then the reduced cost $r(i,j)$ is decreased by 
$2\delta$. This can be corrected by increasing the set weight $\mu(E_l)$ of
exactly one set $E_l$ containing the edge $(i,j)$ by $2\delta$.
Two questions have to be answered.
\begin{enumerate}
\item
What is the accurate edge set $E_l$ for increasing its set weight?
\item
What is the appropriate value $\delta$?
\end{enumerate}
To answer the first question, let us consider MDFS which is used as a subroutine 
during the search step. 
Review the definitions and properties of the sets $L_{[w,A]}$ and $D_{[q,A]}$ as given 
on Pages \pageref{WM1} and \pageref{WM2}, respectively. At the moment when an edge set $E_q$ 
is chosen to obtain a positive set weight $\mu(E_q)$, the set $E_q$ corresponds to the 
current set $D_{[q,A]}$ as defined on Page \pageref{WM2} and $E_q$ enters $\cal F$.
Note that $\cal F$ can contain edge sets which are generated during a previous run
of MDFS. The corresponding set $D_{[q,A]}$ has not to be current with respect to the
current run of MDFS. Hence, we introduce an analogous terminology to ``current'' with
respect to the edge sets in $\cal F$.
A set $E_q \in \cal F$ is called {\em maximal\/} iff $E_q \not\subseteq E_{q'}$ for all
$E_{q'} \in {\cal F} \setminus \{E_q\}$.
Now, it is useful to investigate the structure of a set $D_{[q,A]}$. Let
$$D'_{[q,A]} := \{q\} \cup \{p \in V \mid [p,A] \in D_{[q,A]}\}$$
and
$$\tilde D_{[q,A]} := \{[p,A],[p,B] \mid p \in D'_{[q,A]}\}.$$
The node $q$ is the unique node $p \in D'_{[q,A]}$ such that $p$ is end node of an
edge $(r,p) \in M$ with $r \not\in D'_{[q,A]}$. 
If a path $P$ runs through a set $\tilde{D}_{[q,A]}$ then there is an edge such
that $P$ enters $\tilde{D}_{[q,A]}$ using this edge and also an edge such that $P$ 
leaves $\tilde{D}_{[q,A]}$ using that edge. 
We say that a path $P$ {\em enters\/} or {\em leaves\/} $\tilde D_{[q,A]}$ 
{\em via an edge\/} {\em in\/} $E \setminus M$ if the used edge $([x,B],[y,A])$ 
corresponds to an edge $(x,y) \in E \setminus M$.
Let $(r,q) \in M$ be the unique matched edge with end node $q$. 
During the execution of MDFS, for an $M$-augmenting path $P$, there are three 
possibilities to run through a set $\tilde D_{[q,A]}$.
\begin{enumerate}
\item
$P$ enters $\tilde D_{[q,A]}$ via the matched edge $([r,A],[q,B])$ and leaves
$\tilde D_{[q,A]}$ via an edge in $E \setminus M$.
\item
$P$ enters $\tilde D_{[q,A]}$ via an edge in $E \setminus M$ and leaves 
$\tilde D_{[q,A]}$ via the matched edge $([q,A],[r,B])$.
\item
$P$ enters and leaves $\tilde D_{[q,A]}$ via an edge in $E \setminus M$.
\end{enumerate}
If an $M$-alternating path $R$ enters $\tilde D_{[q,A]}$ via the edge 
$([r,A],[q,B])$ then, by Lemma \ref{Lemma3}, for all $v \in D'_{[q,A]}$, 
$[v,B] \in B_T$. Therefore, with respect to each edge $([i,X],[j,\overline{X}])$ 
in $E^*_M$ with both end nodes in $\tilde{D}_{[q,A]}$, the node weights $\pi(i)$ and
$\pi(j)$ have been decreased by $\delta$ such that we have to increase the set
weight of exactly one edge set containing the edge $(i,j)$ by $2\delta$.
Note that for all $v \in V$ there exists at most one current $D_{[q,A]}$
such that $v \in D'_{[q,A]}$. Hence, we define the edge set $E_q$ 
corresponding to $D'_{[q,A]}$ where $D_{[q,A]}$ current by
$$E_q := (D'_{[q,A]} \times D'_{[q,A]}) \cap E.$$
If we have to increase the set weight with respect to an edge $([i,B],[j,A]) \in E^*_M$ 
then we choose the edge set $E_q$ where $D_{[q,A]}$ is the current set such that 
$i,j \in D'_{[q,A]}$. Hence, for all current sets $D_{[q,A]}$, we perform the following
operation:
\begin{itemize}
\item
If $E_q$ is not already in $\cal F$ then $E_q$ is inserted into $\cal F$ with 
$\mu(E_q) := 2\delta$. Otherwise, $\mu(E_q)$ is increased by $2\delta$. 
\end{itemize}
The following lemma shows that in this situation, the current set $D_{[q,A]}$ always exists.
\begin{lemma} \label{pd1}
Let $[i,B],[j,B] \in B_T$ and $([i,B],[j,A]),([j,B],[i,A]) \in E^*_M$. Then there exists 
a current set $D_{[q,A]}$ such that $i,j \in D'_{[q,A]}$.
\end{lemma} 
{\bf Proof:} 
Note that $([i,B],[j,A]),([j,B],[i,A]) \in E^*_M$ iff $r(i,j) = 0$.
With respect to the positions of the nodes $[i,B]$ and $[j,B]$ in the MDFS-tree $T$, exactly
the following two situations are possible: 
\begin{enumerate}
\item
There is no path $P$ from the root $s$ to a leaf of $T$ such that both
nodes are on $P$.
\item
There is a path $P$ from the root $s$ to a leaf of $T$ such that both
nodes are on $P$.
\end{enumerate}
If the first situation arises then there exists a unique node in $T$ such that both nodes 
$[i,B]$ and $[j,B]$ are successors of this node in $T$ and no successor of this node in $T$ 
has this property. Since no $M$-augmenting path has been found during the last MDFS, this 
node cannot be the node $s$. Hence, this node has to be a node $[u,B] \in V_B$.
Let $s,Q,[u,B]$ be the path from $s$ to the node $[u,B]$ in $T$. Furthermore,
let $[u,B],P_1,[i,B]$ and $[u,B],P_2,[j,B]$ be the paths in $T$ from $[u,B]$ to $[i,B]$ and 
$[j,B]$, respectively. 
By the definition of $L_{[j,A]}$ and $L_{[i,A]}$,
because of the paths $s,Q,[u,B],P_1,[i,B],[j,A],r(P_2),[u,A]$ and
$s,Q,[u,B],P_2,[j,B],[i,A],r(P_1),[u,A]$, $L_{[j,A]} = L_{[i,A]} = [u,A]$ if $[u,A] \not\in A_T$. 
If $[u,A] \in A_T$, i.e., PUSH$([u,A])$ has been performed, then, by Lemma \ref{Lemma4},
$L_{[j,A]} = L_{[i,A]} = L_{[u,A]}$. By the definitions of the current set $D_{[q,A]}$ and
the set $D'_{[q,A]}$, the assertion follows directly.

Assume that the second situation arises. W.l.o.g., let $[j,B]$ be a successor of $[i,B]$ 
in $T$. Let $s,Q,[i,B],R,[j,B]$ be the path from $s$ to $[j,B]$ in $T$. 

If $[i,A] \not\in A_T$ then, because of the path $s,Q,[i,B],[j,A],r(R),[i,A]$ and
the definition of $L_{[j,A]}$, it follows $L_{[j,A]} = [i,A]$. By the definition
of $D_{[i,A]}$, the node $[j,A]$ is contained $D_{[i,A]}$. The definition of $D'_{[i,A]}$ implies 
$i,j \in D'_{[i,A]}$ such that the assertion follows directly.

If $[i,A] \in A_T$ then, by Lemma \ref{Lemma4}, $L_{[j,A]} = L_{[i,A]}$. Hence, the assertion 
follows directly.
\QED

\medskip
Let us examine the effect of the augmentation of the path $P$ to the number of edges
in the current matching with both end nodes in $D'_{[q,A]}$.
If $P$ enters and leaves $\tilde{D}_{[q,A]}$ via an edge in $E \setminus M$, then this number  
decreases by one. In the other two cases, this number does not change.
Hence, the augmentation of an augmenting path which goes through $\tilde{D}_{[q,A]}$
using the unique matched edge with one end node in $D'_{[q,A]}$ and the other end node 
not in $D'_{[q,A]}$ is always allowed. But the augmentation of an augmenting path which
enters and leaves $\tilde{D}_{[q,A]}$ via an edge in $E \setminus M$
is only allowed if $\mu(E_q) = 0$; i.e., $E_q$ is not contained in $\cal{F}$.
Now, we shall determine the accurate value for $\delta$.

Since all node weights have to be non-negative, $\delta$ cannot be larger 
than the node weight of any $M$-free node $i$. Note that with respect to an $M$-free
node $i$, always $[i,B] \in B_T$ is fulfilled. Since all nodes are initialized with the same 
node weight and always the node weights of all nodes in $B_T$ are decreased using the same 
value, all free nodes have the same node weight and no non-free node has a smaller node weight.
Hence, $\delta \leq \pi(i)$ where $i$ is $M$-free implies that after the 
change of the dual values, all node weights are non-negative.

Let $[i,B] \in B_T$. Then $\pi(i)$ will be decreased by $\delta$. For $(i,j) \in E$, 
in dependence of the status of the other end node $j$ with respect to $B_T$ and $A_T$, 
three situations can arise:

If $[j,B] \not\in B_T$ and $[j,A] \not\in A_T$ then $\pi(j)$ will not be changed by the
extension step. Hence, we have to choose $\delta := r(i,j)$ 
to decrease the reduced cost $r(i,j)$ of the edge $(i,j)$ to zero.
 
If $[j,B] \not\in B_T$ and $[j,A] \in A_T$ then $\pi(j)$ will be increased by $\delta$.
Hence, independently of the choice of $\delta$, the reduced cost $r(i,j)$ does not change.

If $[j,B] \in B_T$ then $\pi(j)$ will be also decreased by $\delta$. Hence, we have to 
choose $\delta := \frac{1}{2}r(i,j)$ to decrease $r(i,j)$ to zero.

Note that $\delta$ has to be chosen in such a way that after the extension step
$r(i,j) \geq 0$ for all edges $(i,j) \in E$. Hence, $\delta$ should not be 
larger than the minimal reduced cost with respect to edges $(i,j)$ with
$[i,B] \in B_T$, $[j,B] \not\in B_T$ and $[j,A] \not\in A_T$, and also not larger than 
the half of the minimal reduced cost with respect to edges $(i,j)$ with 
$[i,B],[j,B] \in B_T$, $r(i,j) > 0$, and there is no current $D_{[q,A]}$ such that 
$i,j \in D'_{[q,A]}$. Note that $i,j \in D'_{[q,A]}$ with respect to a current set
$D_{[q,A]}$ implies that $r(i,j)$ does not change since $\mu(E_q)$ is increased by 
$2\delta$. Because of Lemma \ref{pd1}, $r(i,j) = 0$ implies that there is a current 
$D_{[q,A]}$ such that $i,j \in D'_{[q,A]}$.

Since MDFS obtains in addition to the input graph the current set $\cal F$ and the  
corresponding set weights, because of the maintenance of the third optimality 
condition, the following holds true with respect to a maximal $E_q \in \cal F$:

Assume that during the MDFS no path uses the edge $([r,A],[q,B])$ to enter $\tilde{D}_{[q,A]}$; 
i.e., $[q,B] \not\in T_{exp}$. If there is a path $R$ entering $\tilde{D}_{[q,A]}$ via an edge 
in $E \setminus M$ then $\mu(E_q) > 0$ implies that $R$ has to leave $\tilde{D}_{[q,A]}$ via 
the edge $([q,A],[r,B])$, independently if the node $[q,A]$ is already pushed or not. Since
$(q,r) \in M$ and therefore $([q,A],[r,B]) \in E^*_M$, this is always possible.
Note that more than one such a path can run through $\tilde{D}_{[q,A]}$.
Depending on the nodes in $\tilde{D}_{[q,A]}$ which are entering nodes of such a path $R$,
the following situations can happen with respect to a node $v \in D'_{[q,A]}$.
\begin{itemize}
\item[a)]
$[v,B] \in B_T$,
\item[b)]
$[v,B] \not\in B_T$ but $[v,A] \in A_T$, or
\item[c)]
$[v,B] \not\in B_T$ and $[v,A] \not\in A_T$.
\end{itemize}
The problem to solve is the following: How to change the node weights of the
nodes in $D'_{[q,A]}$?
Each entering node $[v,A]$ of $D_{[q,A]}$ is the head of an edge $([x,B],[v,A])$
where $[x,B] \in B_T$. Hence, $\pi(x)$ is decreased by $\delta$. According to 
the first optimality condition,  $\pi(v)$ has to be increased by $\delta$. 
Possibly, there are edges in $E_q$ with exactly one end node is an entering node, 
with both end nodes are entering nodes or with no end node is an entering node. 
With respect to all these cases, the node weights and $\mu(E_q)$ have to be changed 
in such a way that the optimality conditions remain to be valid.
For doing this, we increase $\pi(v)$ by $\delta$ for all $v \in D'_{[q,A]}$.
Since we have increased the reduced cost of all edges in $E_q$ by $2\delta$,
we decrease $\mu(E_q)$ by $2\delta$. Since $\mu(E_q)$ has to be non-negative, 
$\delta$ has to be chosen such that before the change of the dual variables,
$\mu(E_q) \geq 2\delta$. 

Remember that we consider the situation that $[q,B] \not\in T_{exp}$.
Possibly, there exist nodes $[i,B] \in \tilde D_{[q,A]}$ which are
in $T_{exp}$. Since we cannot decrease $\pi(i)$ by $\delta$ and increase $\pi(i)$ by
$\delta$ at the same time, it is not allowed for such a node to be in $B_T$. 
Hence, we define 
$$
B_f := \{[i,B] \mid [i,B] \in \tilde D_{[q,A]}, E_q \mbox{ maximal and }
[q,B] \not\in T_{exp}\}
$$
and we redefine
$$B_T := (V_B \cap T_{exp}) \setminus B_f.$$
Altogether, $\delta$ can be defined in the following way.
\begin{eqnarray*}
\delta_0 & := & \pi(i), \mbox{ where $i$ is $M$-free}, \\
\delta_1 & := & \min\{r(i,j) \mid  [j,A] \not\in A_T, [j,B] \not\in B_T \mbox{ and } 
[i,B] \in B_T \}, \\
\delta_2 & := & \min\{r(i,j)  \mid [i,B],[j,B] \in B_T, r(i,j) > 0 \mbox{ and there is 
no current} \\ 
& & \qquad \qquad \qquad D_{[q,A]} \mbox{ with } i,j \in D'_{[q,A]}\}, \mbox{ and} \\
\delta_3 & := & \min\{\mu(E_q) \mid E_q \in {\cal F} \mbox{ maximal}, [q,B] \not\in B_T 
\mbox{ and } [q,A] \in A_T\}.
\end{eqnarray*}
Then we define
$$\delta := \min \left\{\delta_0, \delta_1, \frac{\delta_2}{2}, \frac{\delta_3}{2}\right\}.$$
Altogether, we have obtained the following extension step.

\begin{tabbing}
AA \= AA \= \kill
$\delta_0 := \pi(i)$ for an $M$-free node $i$; \\
$\delta_1 := \min\{r(i,j) \mid [j,A] \not\in A_T, [j,B] \not\in B_T \mbox{ and } [i,B] \in B_T\}$;\\
$\delta_2 := \min\{r(i,j)  \mid [i,B],[j,B] \in B_T, r(i,j) > 0 \mbox{ and there is no current}$   \\
 $\qquad \qquad \qquad \qquad D_{[q,A]} \mbox{ with } i,j \in D'_{[q,A]}\}$; \\
$\delta_3 := \min\{\mu(E_q) \mid E_q \in {\cal F} \mbox{ maximal}, [q,B] \not\in B_T 
\mbox{ and } [q,A] \in A_T\}$; \\
$\delta := \min \left\{\delta_0, \delta_1, \frac{\delta_2}{2}, \frac{\delta_3}{2}\right\}$; \\
{\bf for} all $[i,B] \in B_T$ \\
\> {\bf do} \\
\> \> $\pi(i) := \pi(i) - \delta$ \\
\> {\bf od}; \\ 
{\bf for} all ($[i,B] \not\in B_T$, $[i,A] \in A_T$) {\bf or} ($i \in D'_{[q,A]}$, $E_q \in \cal F$ 
maximal with \\
\> \> \qquad \qquad \qquad \qquad \qquad \qquad $[q,B] \not\in B_T$ but $[q,A] \in A_T$) \\
\> {\bf do} \\
\> \> $\pi(i) := \pi(i) + \delta$ \\
\> {\bf od}; \\ 
{\bf for} all current $D_{[q,A]}$ and $[q,B] \in B_T$ \\
\> {\bf do} \\
\> \> $\mu(E_q) := \mu(E_q) + 2\delta$ \\
\> {\bf od}; \\
{\bf for} all maximal $E_q \in \cal F$, $[q,B] \not\in B_T$ and $[q,A] \in A_T$ \\
\> {\bf do} \\
\> \> $\mu(E_q) := \mu(E_q) - 2\delta$ \\
\> {\bf od}.
\end{tabbing}

The correctness of the described primal-dual method follows from the 
discussion done during the development of the method. Note that the primal-dual method described 
above is equivalent to the method developed by Edmonds \cite{Ed2}.

\subsection{An implementation of the primal-dual method}

First we shall determine an upper bound for the number of dual changes which can occur between 
two augmentations in the worst case. Then we shall describe the implementation of the search
steps between two augmentations. Finally, we shall describe an implementation of the
computation of $\delta$ and the update of the dual variables. With respect to the determination
of the upper bound, we have to consider four cases.

\medskip
\noindent
{\em Case 1:\/} $\delta = \delta_0$

\medskip
Then after the change of the dual variables, $\pi(i) = 0$ for all $M$-free nodes
$i \in V$. Therefore, the current matching $M$ is of maximum weight and the 
algorithm terminates. Hence, Case 1 occurs at most once.

\medskip
\noindent
{\em Case 2:\/} $\delta = \delta_1$

\medskip
Then during the next search step, at least one new node $[j,A]$ enters $A_T$.
Hence, Case 2 occurs at most $n$ times.

\medskip
\noindent
{\em Case 3:\/} $\delta = \frac{\delta_2}{2}$

\medskip
Then during the next search step, at least one new edge $(i,j)$ enters 
$E^*_M$. Furthermore, at the moment of the definition of the value $\delta$,
there is no current $D_{[q,A]}$ with $i,j \in D'_{[q,A]}$. Lemma \ref{pd1}
shows that after the next search step there exists a current set $D_{[q,A]}$
such that $D'_{[q,A]}$ contains both nodes $i$ and $j$. This means, by the
union of two smaller current sets, a larger current set is obtained. 
Between two augmentations, the number of such unions is bounded by $n-1$.
Hence, Case 3 occurs at most $n-1$ times. 

\medskip
\noindent
{\em Case 4:\/} $\delta = \frac{\delta_3}{2}$

\medskip
Then at least one edge set $E_q$ leaves the family $\cal F$. As long as 
$[q,B] \not\in B_T$, no edge set $E'_q$ can enter the family $\cal F$.
But if $[q,B]$ is pushed, a corresponding edge set $E'_q$ cannot contribute to 
the definition of the value $\delta_3$ before $[q,B]$ leaves $B_T$ again. 
This cannot happen before the next augmentation. Hence, Case 4 occurs at most 
$n$ times.

\medskip
Altogether we have shown that the number of dual changes between two augmentations
is bounded by $3n$.

\medskip
With respect to the implementation of the search steps between two augmentations,
MDFS has to be adapted to the situation that the current search path enters
the set $\tilde{D}_{[q,A]}$ via an edge in $E \setminus M$
with respect to a maximal edge set $E_q \in \cal F$.
According to the discussion during the development of the primal-dual method,
we have to jump to the node $[q,A]$ and to continue the search using the unique edge
$([q,A],[r,B]) \in E^*_M$. This can be organized using a data structure for
disjoint set union. But this data structure has also to support a further operation.

If according to an extension step, $\mu(E_q)$ becomes zero, the corresponding edge
set leaves the family $\cal F$. Note that $E_q$ is a maximal set in $\cal F$. 
If there is a set $E_{q'} \subset E_q$ in $\cal F$ then another set in $\cal F$ becomes 
to be maximal. This means that we have to undo the union operation performed with respect 
to the set $E_q$. 

With respect to the implementation of MDFS, we have introduced a data structure for
disjoint set union which uses the weighted union heuristic. Our goal is now to extend
this data structure to support also deunion operations such that the time used for the 
deunion operations will be, up to
a small constant factor, the same as the time used for the union operations and each 
find operation uses only constant time. This can be done in the following way:

During the execution of an union operation, instead of changing a pointer, we add a
new pointer. The current pointer of an element will be always the last created pointer.
Since the pointer of any element is changed at most $\log n$ times, for each element, 
at most $\log n$ extra pointer are used. The time used for the union operations
remains essentially the same. A deunion can be performed by the deletion of the current 
pointers created during the corresponding union operation, the update of the set sizes 
and the update of the name of the larger subset. The time used for a find operation
remains to be constant. Altogether, the data structure for 
union-find-deunion can be implemented such that the time used for at most $m$ find,
$n$ union and $n$ deunion operations is $O(m + n \log n)$.

After an extension step, the last MDFS can be continued instead of
to start a new MDFS. 
Next we shall describe an implementation of the computation of the value $\delta$ and
the update of the dual variables.

\medskip
Since all $M$-free nodes $i$ have the same dual weight, $\delta_0$
can be computed by the consideration of any $M$-free node in constant time.

For the computation of $\delta_1$, we maintain a priority queue ${\cal P}_1$
which contains for all $[j,A] \not\in A_T$ with $[j,B] \not\in B_T$ an item
which points to a list containing exactly the edges $(i,j)$ with $[i,B] \in B_T$
of minimum reduced cost within all such edges, if such an edge exists. The 
associated key with this item is the minimum reduced cost of these edges.
Furthermore, using an array of size $n$, we have direct access to the list in ${\cal P}_1$ 
containing all edges with end node $[j,A]$, if such an edge in ${\cal P}_1$ exists.
Note that each extension step decreases the keys of all items in the
priority queue by the current $\delta$. It is useful to maintain the property
that always the key of all elements in ${\cal P}_1$ has to be decreased by the same
amount. Hence, we maintain the sum $\Delta_1$ of all dual changes
done so far. If we add an edge $(i,j)$ to the priority queue then we define
the modified reduced cost of the edge to be $r(i,j) + \Delta_1$.
At the moment when a node $[i,B]$ enters $B_T$, it is possible that we have to insert some
edges into ${\cal P}_1$. Hence, we update ${\cal P}_1$ with respect to $[i,B] \in B_T$ 
at that moment in the following way:

\medskip
For all edges $(i,j)$ with $[j,B] \not\in B_T$ and $[j,A] \not\in A_T$ perform 
the following update operations: 
\begin{itemize}
\item[(1)]
If no element with respect to $j$ is contained in the priority queue then
we create a list which contains the element $(i,j)$ with modified reduced cost 
$r(i,j) + \Delta_1$. We insert a new item which points to this list with associated
key $r(i,j) + \Delta_1$ into ${\cal P}_1$.
\item[(2)]
If ${\cal P}_1$ contains an item which points to a list containing edges with end node
$[j,A]$ with larger associated key than $r(i,j) + \Delta_1$ 
then delete all edges from the list, insert the edge by $([i,B],[j,A]))$ into the list and 
decrease the key of the item such that its value becomes $r(i,j) + \Delta_1$.
\item[(3)]
If ${\cal P}_1$ contains an item which points to a list containing edges with end node
$[j,A]$ with associated key $r(i,j) + \Delta_1$ then add the edge $([i,B],[j,A]))$ to the 
list.
\item[(4)]
In all other cases do nothing.
\item[(5)]
If ${\cal P}_1$ contains some edges with respect to the node $[i,A]$ then 
delete the corresponding item and the list of edges to which this item points.
\end{itemize}
The number of deletions performed in Step 5 is bounded by the number of nodes
in $V_A$. If $\delta = \delta_1$, we have to delete at least one minimum
key from ${\cal P}_1$. Altogether, the total number of deletions is bounded by $n$.
For the implementation of the priority queue ${\cal P}_1$ we can use
Fibonacci heaps \cite{FrTa} or strict Fibonacci heaps \cite{BrLaTa} such that
with respect to the computation of all $\delta_1$'s between two
augmentations, the used time is $O(m + n \log n)$.

For the computation of all $\delta_2$'s, we maintain a priority queue 
${\cal P}_2$ which contains all edges $(i,j)$ such that $[i,B], [j,B] \in B_T$ 
and $r(i,j) > 0$.
Again, we can use Fibonacci heaps or strict Fibonacci heaps for the realization 
of the priority queue.
Note that each extension step decreases all weights of the elements $(i,j)$ 
in ${\cal P}_2$ with the property that there is no current $D_{[q,A]}$ such that 
$i,j \in D'_{[q.A]}$ by $2\delta$ where $\delta$ is the value chosen for the 
current extension step.
Analogously to the manipulation of ${\cal P}_1$, we maintain with respect to 
${\cal P}_2$ the sum $\Delta_2$ of all dual changes done so far with respect to 
edges in ${\cal P}_2$ and modify the weights in the appropriate manner.
We update ${\cal P}_2$ with respect to $[i,B] \in B_T$ 
at the moment when $[i,B]$ is added to $B_T$ in the following way:
\begin{itemize}
\item
For all $[j,B] \in B_T$ with $r(i,j) > 0$ add the edge $(i,j)$ with modified
reduced cost $r(i,j) + \Delta_2$ to the priority queue ${\cal P}_2$.
\end{itemize}
Since at most $m$ edges are inserted, the used time is $O(m)$.
 
For the computation of $\delta_2$, we have to find the smallest element
$(i,j)$ in ${\cal P}_2$ which has the property that there exists no current $D_{[q,A]}$
such that $i,j \in D'_{[q,A]}$. We use the priority queue ${\cal P}_2$ and
perform the following procedure:

\begin{tabbing}
AA \= AA \= \kill
(1) findmin; \\ 
\> (* Let $(i,j)$ be the output of findmin.*) \\
(2) {\bf if} there exists $D_{[q,A]}$ current with $i,j \in D'_{[q,A]}$ \\
\> {\bf then} \\
\> \> deletemin; \\
\> \> {\bf goto} (1) \\
\> {\bf else} \\ 
\> \> $\delta_2 := r(i,j) - \delta_2$ \\
\> {\bf fi}.
\end{tabbing}

If $\delta = \delta_2$ then we have to delete at least one minimal
element from the priority queue ${\cal P}_2$. The number of deletemin operations
performed during all computations of $\delta_2$ between two augmentations is bounded by 
the number of edges in $E$.
Altogether, the number of deletions performed during the computation of
$\delta_2$ is $O(m)$. Each deletion can be performed in $O(\log n)$ time.
Hence, the total time used for deletions is $O(m \log n)$.
Altogether, with respect to the computation of all $\delta_2$'s between two
augmentations, the used time is $O(m \log n)$.

For the computation of all $\delta_3$'s, we maintain a priority queue 
${\cal P}_3$ which contains for all maximal $E_q \in \cal F$ with $[q,B] 
\not\in B_T$ and $[q,A] \in A_T$ the value $\mu(E_q)$.
We use a heap, a Fibonacci heap or a strict Fibonacci heap for the realization of 
the priority queue ${\cal P}_3$.
Each extension step decreases all keys of the elements in $P_3$. The amount
is two times the current $\delta$. Hence, we can use the value $\Delta_2$
defined above and modify the keys in the appropriate manner.
We update ${\cal P}_3$ before the computation of $\delta$. We have to insert 
for all $[q,A] \in V_A$ such that $E_q$ is maximal and $[q,A]$ was inserted 
into $T_{exp}$ after the last dual change and the last augmentation, respectively 
but $[q,B] \not\in T_{exp}$ the value $\mu(E_q) + \Delta_2$. 
We have to delete for all $[q,B]$ which are inserted to $T_{exp}$
after the last dual change and the last augmentation, respectively 
the corresponding value if in ${\cal P}_3$ such a value exists. 
Since the number of dual changes between two augmentations is bounded by $3n$ and
$[q,B] \in T_{exp}$ can only leave $T_{exp}$ because of an augmentation, the number
of insertions and deletions is at most $2n$. 
If $\delta = \delta_3$, we have to delete at least one minimal
element from ${\cal P}_3$. As observed above, the number of such deletions is
bounded by $n$. If there exists some sets $E_{q'} \subset E_q$ in $\cal F$,
some other edge sets in $\cal F$ become to be maximal. Hence, we have to perform
the deunion operation which corresponds to the union operation performed with
respect to the construction of $E_q$ and to insert the resulting maximal sets $E_{q'}$ of
$\cal F$ with key $\mu(E_{q'}) + \Delta_2$. As observed above, the total number of such 
insertions is also bounded by $n$. 
Each deletion and each insertion can be performed in $O(\log n)$ time.
Altogether, with respect to the computation of all $\delta_3$'s between two
augmentations, the used time is $O(n \log n)$.

We have proved the following theorem.
\begin{theo}
The primal-dual method can be implemented such that its time complexity is $O(nm \log n)$.
\end{theo}
To get an implementation of time complexity $O(nm \log n)$, we can use ordinary heaps
instead of Fibonacci heaps or strict Fibonacci heaps. Using Fibonacci heaps or strict
Fibonacci heaps, the value $\delta_1$ can be computed such that the used time between
two augmentations is $O(m + n \log n)$. To get an implementation of time complexity
$O(n(m + n \log n))$ the only critical point is the computation of the values $\delta_2$
such that the time used between two augmentations is bounded by $O(m + n \log n)$.
Gabow \cite{Ga2} has presented an implementation of Edmonds maximum weighted matching 
algorithm which uses complicated data structures at the first SODA. The stated time 
complexity is $O(n(m + n \log n))$.

The description of some recent programs which implements Edmonds' maximum weighted 
matching algorithm can be found in \cite{CoRo, MeSc, Ko}.

\medskip
\noindent
{\bf Acknowledgments:} 
I thank Ross McConnell for pointing out a mistake in the description of the
algorithm in \cite{Bl3} and Ari Freund for indicating some points in 
\cite{Bl2} which have to be clarified. These hints
have caused myself to revise my whole work on graph matching algorithms.
I thank Mathias Hauptmann for a careful reading of a preliminary version of the paper.

\end{document}